\documentclass[preprint,aps]{revtex4}
\pdfoutput=1
\usepackage{dcolumn}% Align table columns on decimal point
\usepackage{bm}% bold math
\usepackage{graphicx}% Include figure files
\usepackage{epsfig}
\usepackage{amsmath}
\usepackage{amssymb}
\usepackage{mathrsfs}
\usepackage{ulem}
\usepackage{color}
\usepackage{hyperref}
\usepackage{float}
\usepackage{braket}
\allowdisplaybreaks

\def\fun#1#2{\lower3.6pt\vbox{\baselineskip0pt\lineskip.9pt
  \ialign{$\mathsurround=0pt#1\hfil##\hfil$\crcr#2\crcr\sim\crcr}}}

\def\simlt{\stackrel{<}{{}_\sim}}
\def\simgt{\stackrel{>}{{}_\sim}}

\input epsf

\newenvironment{Eqnarray}%
         {\arraycolsep 0.14em\begin{eqnarray}}{\end{eqnarray}}
\newcommand{\be}{\begin{equation}}
\newcommand{\ee}{\end{equation}}
\newcommand{\bea}{\begin{Eqnarray}}
\newcommand{\eea}{\end{Eqnarray}}

\begin{document}

\begin{flushright}
\vspace{-0.2cm}EFI-19-11\\
\end{flushright}

\title{ Dynamical Higgs Field Alignment in the NMSSM }

\vspace*{0.2cm}

\author{
Nina M. Coyle$^a$ and Carlos E.M. Wagner$^{a,b,c}$
\vspace{0.2cm}
\mbox{}
 }
\affiliation{
\vspace*{.5cm}
$^a$~\mbox{Physics Department and Enrico Fermi Institute, University of Chicago, Chicago, IL 60637}\\
$^b$  \mbox{Kavli Institute for Cosmological Physics, University of Chicago, Chicago, IL 60637}\\
$^c$ \mbox{High Energy Physics Division, Argonne National Laboratory, Argonne, IL 60439}\\
}

\begin{abstract}
Experimental probes of the recently discovered Higgs boson show that its behavior is close to that of the Standard Model (SM) 
Higgs particle. Extensions of the SM which include extra Higgs bosons are constrained by these observations, implying either the decoupling of 
the heavy  non-standard Higgs particles or the realization of alignment, associated with vanishing mixing of the SM-like
Higgs boson with the non-standard ones.  Quite generally, alignment is not enforced by  symmetry considerations and
hence it is interesting to look for dynamical ways in which this condition can be realized. We show that this is possible in the
Next-to-Minimal Supersymmetric Standard Model (NMSSM), in which alignment is achieved for values of the coupling of
the Higgs fields to the singlet field that become large close to the Grand Unification (GUT) scale. This, in turn, can be explained
by the composite nature of the Higgs fields, with a compositeness scale close to the GUT scale.  In this article we present this dynamical scenario
and discuss its phenomenological properties. 
\end{abstract}
\thispagestyle{empty}

\maketitle

\section{Introduction}

With the discovery of the Higgs boson in 2012, the Standard Model (SM) is complete and the theory of electroweak symmetry breaking confirmed~\cite{Aad:2012tfa,Chatrchyan:2012xdj}. The primary focus of the Large Hadron Collider (LHC) since this discovery has been measurements of the precise properties of the Higgs boson~\cite{Aad:2015zhl},
 as well as searches for new physics. However, no evidence of new physics beyond the SM has yet been found, and the LHC Higgs boson so far appears to be SM-like.

In light of these results, extensions of the SM have become further constrained, and an interesting area of study is the examination of how extended Higgs theories may include a SM-like Higgs boson~\cite{Gunion:2002zf}.  
This can be achieved
in two ways: either by decoupling of the non-standard physics, rendering the SM as the effective low energy theory, or by the condition of alignment, associated with the cancellation of the mixing of the non-standard
Higgs bosons with the SM-like one.  The condition of alignment has been studied in several extensions of the SM, including two Higgs doublet models, the Minimal Supersymmetric Standard Model extension (MSSM) and the next-to-minimal one (NMSSM)~\cite{Gunion:2002zf,Delgado:2013zfa,Craig:2013hca,Carena:2013ooa,Carena:2014nza,Carena:2015moc,Bernon:2015qea,Haber:2017erd}. 
While the necessary parameter spaces have been identified and studied in the past, of further interest is the manner through which these parameter spaces may be obtained.  Although in certain cases
alignment may be associated with symmetry properties~\cite{Dev:2014yca,Dev:2017org,Benakli:2018vqz,Benakli:2018vjk,Darvishi:2019ltl},  
this is not the case in most extensions of the SM. It is therefore of interest to study whether alignment could be achieved dynamically. 

In this paper, we focus on the NMSSM and investigate how one may dynamically obtain Higgs alignment in this theory. We concentrate on the running of the NMSSM parameters up to the Grand Unification (GUT) scale, and examine general implementations of the high-energy theories suggested by such running. Particular focus is placed on Fat Higgs models, which we show may naturally include the elements necessary to satisfy the alignment limit for the doublet sector as well as limited mixing with the singlet.  For this to happen, the  compositeness scale must be close to the GUT scale. We therefore also examine the interesting coincidence of bottom and tau Yukawa unification at the GUT scale, which may be realized for the same parameter space as the one associated with Higgs alignment.  We also consider the possibility of including a well behaved Dark Matter candidate within this scenario. 

This paper is structured as follows. In Section \ref{sec:review}, we review the alignment limit of the NMSSM and the relevant conditions on the parameters necessary for alignment. In Section \ref{sec:alignment}, we present results from the running of the NMSSM parameters and examine the range of GUT-scale parameter values for which alignment is obtained in the doublet and singlet sectors. We then present an implementation of a Fat Higgs theory which runs down to alignment at the weak scale in Section \ref{sec:FatHiggs}. In Section \ref{sec:Unification}, we examine the bottom- and tau-Yukawa unification for our set of low-energy parameters. Finally, in Section \ref{sec:Conclusion} we present our conclusions. 

\section{The alignment limit of the NMSSM} \label{sec:review}

Within the NMSSM Higgs sector, which contains two doublets and a singlet, there are two methods through which one may obtain a SM-like Higgs of 125 GeV: decoupling and alignment. In the decoupling case, the heavier non-standard Higgs bosons are pushed to high masses, such that the mixing with the SM-like Higgs boson is suppressed. In the case of alignment, the parameters of the Higgs sector are such that the mixing terms of the squared-mass matrix between the SM-like Higgs boson and the neutral, non-SM-like one and singlet are small. More specifically, if we work in the Higgs basis~\cite{higgsbasis,Branco:1999fs} in which only one of the doublets acquires a vacuum expectation value and hence is aligned with the SM Higgs doublet, here denoted by the subscript 1, the symmetric CP-even Higgs mass-squared matrix is given generally by
\begin{equation} 
\mathcal{M}^2 = \begin{pmatrix} \mathcal{M}^2_{11} && \mathcal{M}^2_{12} && \mathcal{M}^2_{13} \\ && \mathcal{M}^2_{22} && \mathcal{M}^2_{23} \\ && && \mathcal{M}^2_{33} \end{pmatrix} 
\end{equation}
and the alignment condition is
\begin{equation}
\mathcal{M}^2_{12},\mathcal{M}^2_{13} \ll \mathcal{O}(v^2).
\end{equation}
With minimal mixing, we also therefore have that
\begin{equation} m_{h}^2 \approx \mathcal{M}^2_{11} = (125 \text{ GeV})^2. \end{equation}
The alignment limit of the NMSSM and its phenomenological properties have previously been thoroughly investigated in Ref.~\cite{Carena:2015moc}. Here we give a brief review of the relevant properties.

We define the relevant couplings defining the interaction of the Higgs fields through the superpotential
\begin{equation}
W = \lambda S H_u H_d +\frac{\kappa}{3} S^3 + h_u  Q H_u U^c_R + h_d H_d Q D^c_R,
\end{equation}
where the Higgsino mass parameter is proportional to the vacuum expectation value of the singlet field  $\mu = \lambda v_s$.
We shall follow the conventions of Refs.~\cite{Carena:2015moc},\cite{Ellwanger:2009dp}. %These are references 9 and 36. Should 36 be moved upward to 13? (references 11,12 are in the first par of the section)

In the Higgs basis $\{H^{SM}, H^{NSM}, H^{S}\}$, where $H^{SM}$ denotes the SM-like Higgs, $H^{NSM}$ the non-standard Higgs doublet contribution and $H^{S}$ the singlet contribution,
the CP-even Higgs tree-level squared-mass matrix can be explicitly written as
\begin{equation}
\begin{pmatrix} 
	\bar{M}_Z^2 c_{2\beta}^2 + \frac{1}{2}\lambda^2v^2 && -\bar{M}_Z^2 s_{2\beta} c_{2\beta} && \sqrt{2} \lambda v \mu (1 - \frac{M_A^2}{4\mu^2} s_{2\beta}^2 - \frac{\kappa}{2\lambda} s_{2\beta}) \\
	 && M_A^2 + \bar{M}_Z^2 s_{2\beta}^2 && -\frac{1}{\sqrt{2}} \lambda v \mu c_{2\beta} ( \frac{M_A^2}{2\mu^2} s_{2\beta} + \frac{\kappa}{\lambda} ) \\
	 && && \frac{1}{4} \lambda^2 v^2 s_{2\beta} (\frac{M_A^2}{2\mu^2} s_{2\beta} - \frac{\kappa}{\lambda}) + \frac{\kappa\mu}{\lambda} ( A_{\kappa} + \frac{4\kappa\mu}{\lambda})
\end{pmatrix}
\end{equation}
where $s_{2\beta} = \sin2\beta$, etc. and we have defined
\begin{equation}
\bar{M}_Z^2 \equiv m_Z^2 - \frac{1}{2} \lambda^2 v^2.
\end{equation}

Including up to the first order stop loop corrections~\cite{mssmhiggsradcorr}--\cite{Lee:2015uza}, the entries involving the doublets are given by
\begin{align}
\mathcal{M}^2_{11} =& \bar{M}_Z^2 c_{2\beta}^2 + \frac{1}{2}\lambda^2 v^2 + \frac{3 v^2 s_{\beta}^4 h_t^4}{8\pi^2} \left[ \ln \left( \frac{M_S^2}{m_t^2} \right) + \frac{X_t}{M_S^2} \left( 1-\frac{X_t^2}{12 M_S^2} \right) \right] \\
\mathcal{M}^2_{22} =& M_A^2 + s_{2\beta}^2 \left( \bar{M}_Z^2 + \frac{3 v^2 h_t^4}{32\pi^2} \left[ \ln \left( \frac{M_S^2}{m_t^2} \right) + \frac{X_t Y_t}{M_S^2} \left( 1 - \frac{X_t Y_t}{12M_S^2} \right) \right] \right) \\
\mathcal{M}^2_{12} =& -s_{2\beta} \left( \bar{M}_Z^2 c_{2\beta} - \frac{3 v^2 s_{\beta}^2 h_t^2}{16\pi^2} \left[ \ln \left( \frac{M_S^2}{m_t^2} \right) + \frac{X_t (X_t + Y_t)}{2M_S^2} - \frac{X_t^3 Y_t}{12 M_S^4} \right] \right)
\label{eq:Higgsmass}
\end{align}
where $X_t = A_t - \mu \cot\beta$, $Y_t = A_t + \mu \tan\beta$, $A_t$ is the stop mixing mass parameter and $M_S$ is the geometric mean of the two stop mass eigenstates.

One may rewrite the expression for $\mathcal{M}^2_{12}$ in terms of $\mathcal{M}^2_{11}$ by relating the first-order stop loop correction terms, in which case the conditions for exact alignment up to first-order stop loop corrections become
\begin{align}
\mathcal{M}^2_{12} &= \frac{1}{\tan\beta} \left[ \mathcal{M}^2_{11} - c_{2\beta} m_Z^2 - \lambda^2 v^2 s_{\beta}^2 \right] + \frac{3 v^2 s_{\beta}^2 h_t^4 \mu X_t}{16\pi^2 M_S^2} \left( 1 - \frac{X_t^2}{6M_S^2} \right) = 0, \label{align12} \\
\mathcal{M}^2_{13} &= \sqrt{2}\lambda v \mu \left( 1 - \frac{M_A^2 s_{2\beta}^2}{4\mu^2} - \frac{\kappa s_{2\beta}}{2\lambda} \right) = 0
\end{align}
Values of the $\mu$ parameter close to the weak scale and therefore much lower than the stop masses are preferred in order to obtain a mostly Bino or singlino Dark Matter (DM) candidate and to reduce
the fine tuning associated with electroweak symmetry breaking~\cite{Baum:2017enm,Brust:2011tb}.  
As shown in Eq.~(\ref{align12}), the stop loop corrections to $\mathcal{M}^2_{12}$ not included in $\mathcal{M}_{11}^2$
are suppressed by $\mu/M_S \ll 1$, and one may therefore neglect the stop corrections to find an approximate relation between the values of $\lambda$ and $\tan\beta$ which satisfy exact alignment. Taking $\mathcal{M}^2_{11} = m_h^2$, Eq.~(\ref{align12}) gives~\cite{Carena:2015moc}
\begin{equation} \label{lambda_align}
(\lambda^{A})^2 = \frac{ m_h^2 - m_Z^2 c_{2\beta} }{v^2 s_{\beta}^2}.
\end{equation}

\begin{figure}[H]
	\centering
	\includegraphics[width=0.75\textwidth]{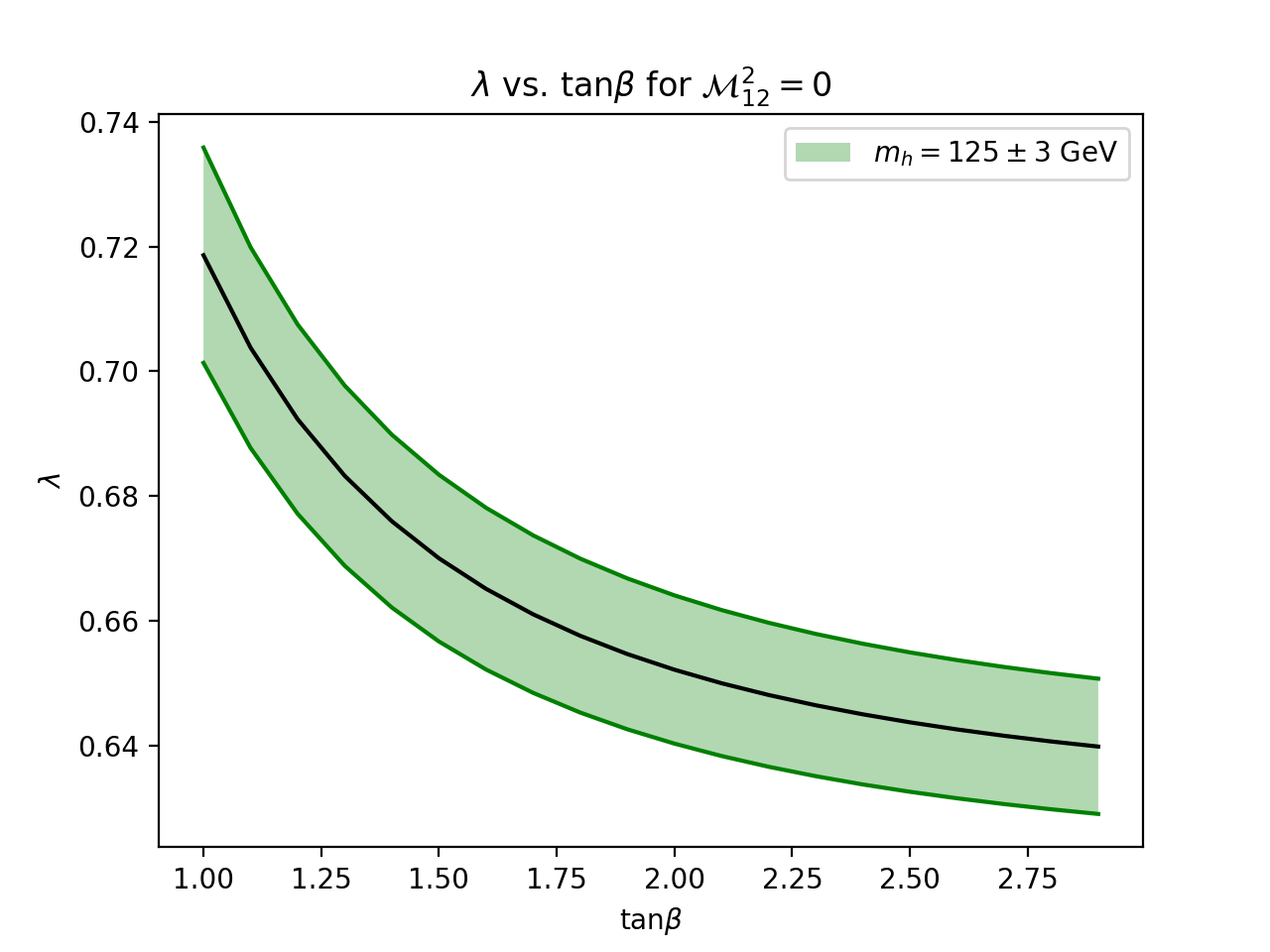}
	\caption{$\lambda$ vs. $\tan\beta$ curves which gives $\mathcal{M}^2_{12}=0$. The solid black line shows exact alignment for $m_h = 125$ GeV. The shaded region covers $m_h = 125 \pm 3$ GeV, with the upper edge corresponding to $m_h = 128$ GeV and the lower edge to $m_h=122$ GeV. }
	\label{exact_align}
\end{figure}

Fig.~\ref{exact_align} shows the $\lambda$ vs. $\tan\beta$ curves given by Eq.~(\ref{lambda_align}) for $m_h = (125 \pm 3) \text{ GeV}$, where we have included an uncertainty of 3~GeV characterizing the theoretical uncertainties in the determination of the Higgs mass. Points within this region will be close to fulfilling exact alignment, while points close to this region should have small mixing between the two doublets. We will better define ``small" mixing quantitatively in our later analyses.  In order to analyze a possible dynamical origin of these parameters, we are interested in identifying the high energy-scale values of NMSSM parameters which naturally run down to this alignment limit at low energies.

Although the above conditions of alignment have been derived by performing an analysis by including only one loop corrections,  models that lead to an appropriate phenomenology at low energies tend 
to be consistent with those conditions, as shown by the similarity of the phenomenological properties of the benchmark scenarios derived in Ref.~\cite{Carena:2015moc} compared with more complete numerical analysis
as those performed in Refs.~~\cite{Ellwanger:2011aa}--\cite{King:2014xwa}.

\section{Running of NMSSM couplings to alignment} \label{sec:alignment}

\subsection{Results of running GUT-scale parameters to weak scale} \label{sec:running}

\begin{figure}[H]
	\centering
	\includegraphics[width=0.75\textwidth]{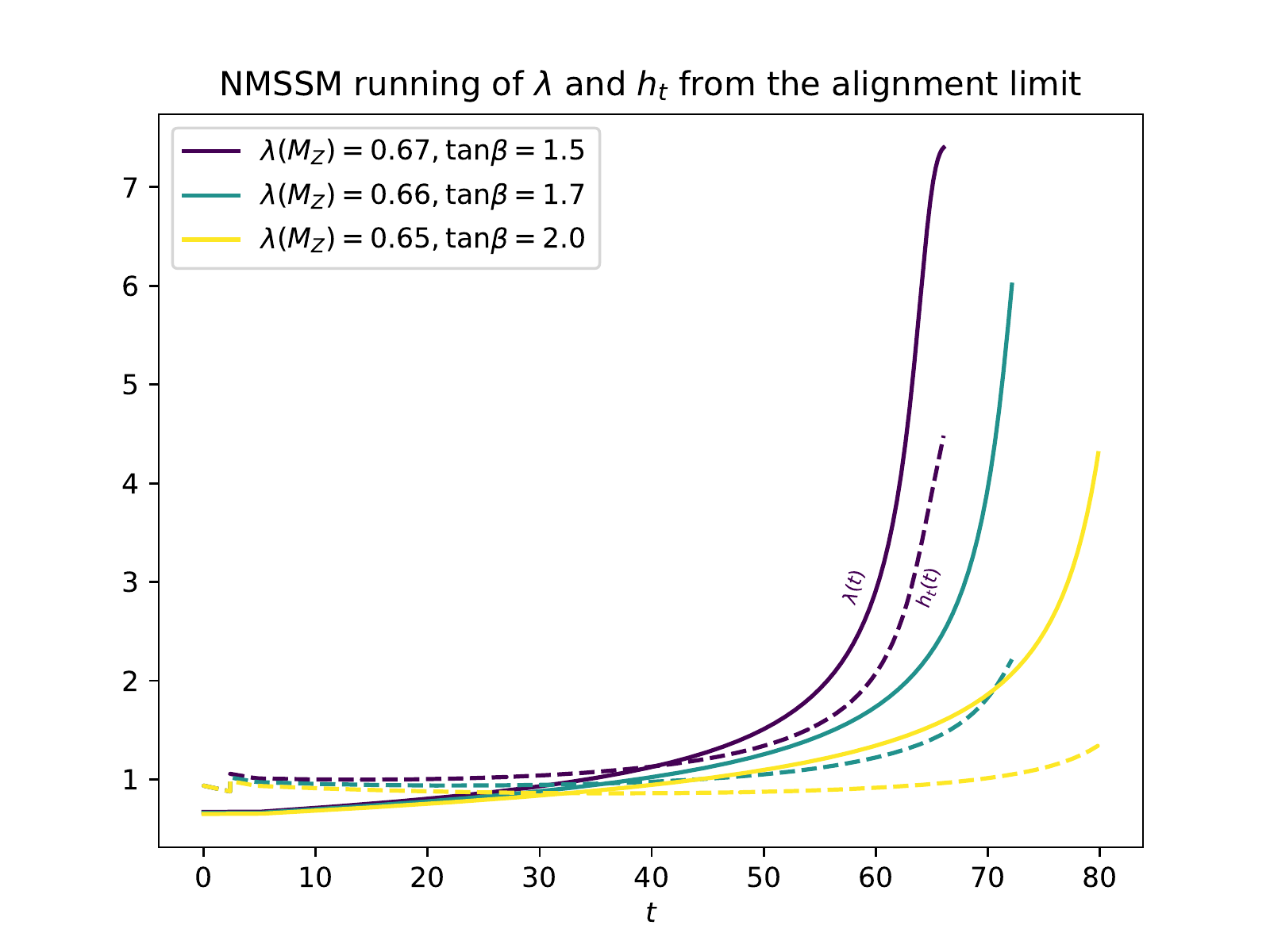}
	\caption{Running of $\lambda$ (solid lines) and $h_t$ (dashed lines) from the 
	weak scale to higher energies, with $t = \ln\left(\frac{Q^2}{M_Z^2}\right)$.  
	We display the running for initial values of $(\lambda(M_Z), \tan\beta) = (0.67, 1.5), (0.66, 1.7),$ and $(0.65,2.0)$, which lie within the alignment region shown in the previous section. 
	%The jump in $h_t$ at the heavy Higgs boson mass scale arises from the boundary condition between the effective theory below $M_{A}$ and the SUSY theory above $M_{A}$.
	}
	\label{RGE}
\end{figure}

As is well known, in minimal low energy supersymmetric models the values of the gauge couplings tend to evolve at a common value at a large energy scale denoted as the Grand 
Unification scale,
$M_{GUT}$, of about 2~$\times \ 10^{16}$~GeV~\cite{Ellis:1990wk,Langacker:1991an,Amaldi:1991cn}.  The values of $\lambda$ and $\tan\beta$ shown in the previous section naturally lead to large values of $\lambda(M_{GUT})$ and $h_{t}(M_{GUT})$ under the NMSSM 
Renormalization Group  equations (RGE)~\cite{Ellwanger:2009dp}. This running seems to suggest a composite nature for the Higgs bosons, for which the relevant couplings, in this case $\lambda$ and $h_t$, become large near the compositeness scale. In particular, if the compositeness scale is of the order of $M_{GUT}$, it appears that one naturally obtains the NMSSM alignment condition ${\cal{M}}_{12}^2=0$ at the weak scale. Fig.~\ref{RGE} shows the general behavior of the running of $\lambda$ and $h_t$ up to the GUT scale. In this plot, we have chosen three different points within the exact alignment region, with a low value of the non-standard Higgs bosons masses, $M_{A} = 300$ GeV and a characteristic stop mass scale $M_{SUSY}=1$~TeV.  Since ignoring decoupling effects $h_t(m_t) \sim m_t(m_t)/(v s_\beta)$,  where $m_t$ is the running top quark mass, the value of $h_t$ becomes weaker at larger values of 
$\tan\beta$.  On the other hand, taking into account decoupling effects, increases in the heavy Higgs boson scale tend to lead to somewhat lower values of $h_t$ at the GUT-scale.

In order to thoroughly examine the range of GUT-scale parameter values for which one obtains Higgs alignment, and to identify the stability of this running to the alignment limit, we begin with a range of values for $\lambda(M_{GUT})$ and $h_t(M_{GUT})$ and run each pair downward in energy. There are three primary regions between $M_{Z}$ and $M_{GUT}$: the low-energy effective SM theory below $M_{A}$, the 2HDM region between approximately $M_A$ and $M_{SUSY}$, and the NMSSM region above $M_{SUSY}$. We employ the relevant RGE equations within each region; the equations for each region are listed in Appendix \ref{sec:RGE}. At the boundary between the SM and 2HDM running at $M_{A}$, we relate the single effective Higgs field in the SM to the two Higgs doublets by $\phi = H_d \cos\beta + i \tau H_u^{*} \sin\beta$. This relation gives 
$h_t^{\rm eff} = h_t \sin\beta$. We approximately identify the scale of the singlet with $M_{SUSY}$, and therefore run the parameter $\lambda$ between $M_{GUT}$ and $M_{SUSY}$, stopping its running below this scale. The value of $\tan\beta$ is determined by requiring that the running top mass is equal to approximately $m_t(M_t)\simeq163$~GeV, leading to a pole top quark mass of approximately the observed value, $M_t \simeq 173$~GeV.

\begin{figure}[H] 
	\centering
	\includegraphics[width=0.75\textwidth]{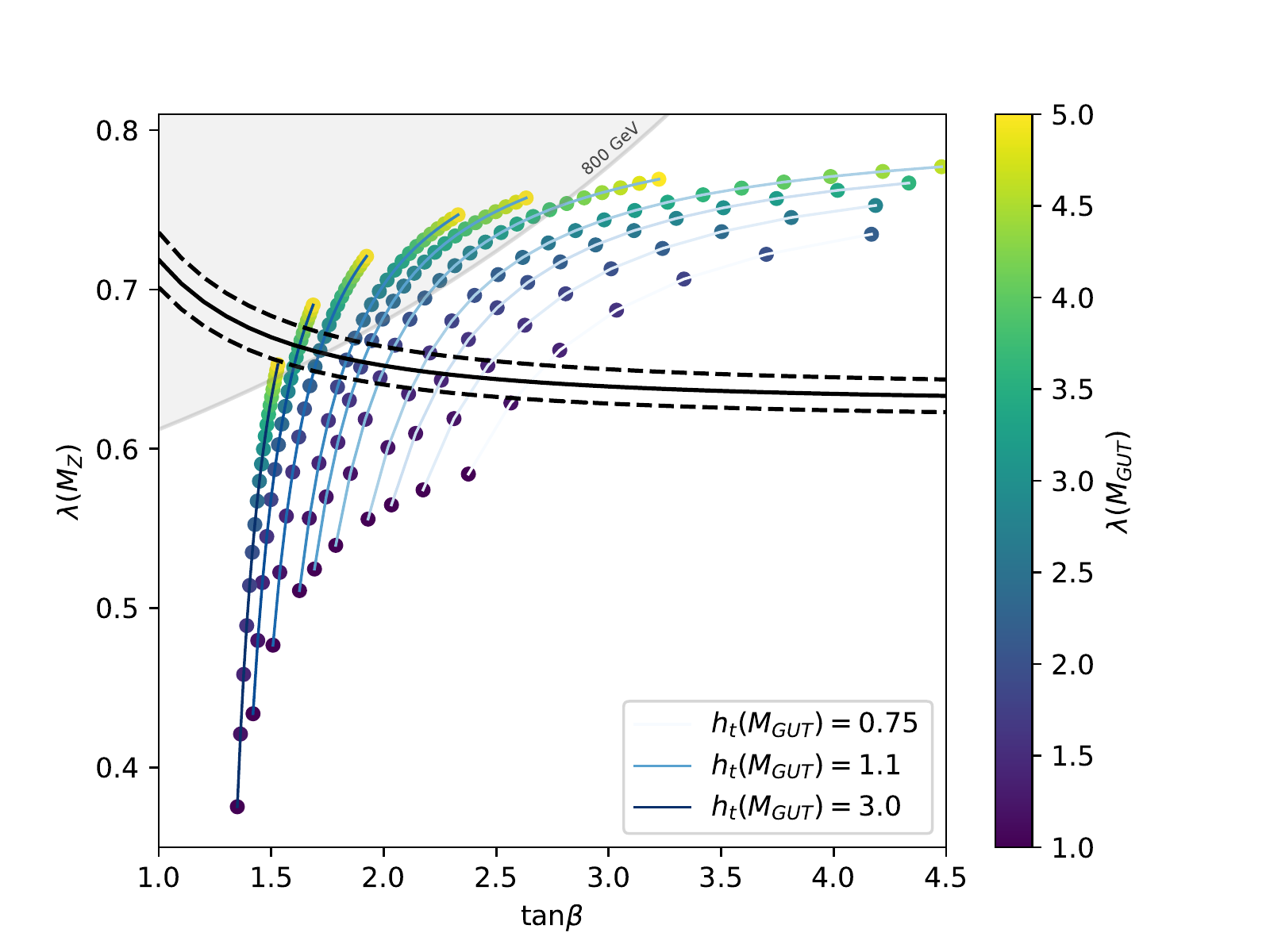} 
	\caption{Plot showing the ($\tan\beta$, $\lambda(M_Z)$) points obtained by running down from $M_{GUT}$ with large $\lambda(M_{GUT})$ and moderate $h_t(M_{GUT})$. The different contours arise from varying $h_t(M_{GUT})$, while the colorbar indicates the value of $\lambda(M_{GUT})$. Results are displayed for $M_{SUSY} = 1$ TeV. The solid and dashed black lines indicate the region of exact alignment for $m_h = 125 \pm 3 \text{ GeV}$. The shaded grey region indicates the region in which it is difficult to obtain a lighter Higgs mass of 125 GeV without tension with existing stop mass limits.}
	\label{contour_plots}
\end{figure}

Fig.~\ref{contour_plots} shows the results of running down from $M_{GUT}$ to $M_Z$, with initial values of $\lambda$ between 1 and 5 and values of $h_t$ between 0.75 and 3.0 at the GUT scale. The value of $\kappa$ is set to 0. We find that the results are stable under TeV-scale variations in the value of the running boundary $M_{SUSY}$, and thus ignore the small thresholds arising from the decoupling of the supersymmetric particles. We display the results for $M_{SUSY}=1$ TeV. The value of $M_A$ is chosen to be 300 GeV.  Significantly larger values of $M_A$, on the order of $M_{SUSY}$, push the $h_t(M_{GUT}) \leq 1$ curves toward large values of $\tan\beta$. For values of $M_A \lesssim 500$ GeV, the results have little variation. 

The obtention of $m_h = 125$ GeV comes into tension with existing stop mass constraints for small values of $\tan\beta$ and large values of $\lambda(M_Z)$, for which the tree level contribution to $m_h$ becomes large. Tree-level contributions close to the observed Higgs mass result in the need for small stop loop corrections and hence small values of the stop masses (see Eq.~\ref{eq:Higgsmass}).  Based on recent results from stop searches \cite{Aaboud:2017ayj, Aaboud:2017aeu, CMS:2019qkm, ATLAS:2019oho}, we use a stop mass bound of $M_S > 800$ GeV.
We employ a lower bound than some of the quoted values after noting that the bounds may be relaxed depending on the specific stop decay paths and neutralino and chargino masses within the model. The scenario presented in Fig. 12 of Ref.~\cite{Aaboud:2017ayj} most closely aligns with the neutralino/chargino spectrum we obtain in scenarios with a realistic Dark Matter candidate, which are further discussed in Section~\ref{sec:Phenomenology}. Splittings between the right- and left-handed stops, multiple decay modes mediated by neutralinos and charginos, and decays through heavier Higgsinos may further relax the 800 GeV bound. In particular, we note that for lightest neutralino masses of order $m_{\tilde{\chi}_{1}^{0}} \gtrsim 200$ GeV the bounds may be relaxed significantly, and in fact no meaningful bounds are placed for $m_{\tilde{\chi}_{1}^{0}} \gtrsim 300$ GeV in that particular analysis.

Moreover, for small stop mixing, a bound on $M_S$
is approximately equivalent from the point of view of the radiative corrections to the Higgs mass to a bound on the
geometric average of the two stop masses. Hence, when comparing with experimental results one should recall that
a bound $M_S > 800$~GeV is approximately equivalent to a bound on the right handed stop mass $m_{\tilde{t}_R} > 600$~GeV and on the left handed stop mass  (which is close in mass to the left handed sbottom)
$m_{\tilde{t}_L} > 1.1$~TeV.  

From the results in Fig.~\ref{contour_plots}, we see that lower values of $h_t$ at the GUT scale tend to push $\tan\beta$ and $\lambda(M_Z)$ to larger values, while lower values of $\lambda(M_{GUT})$ lead to lower values of $\lambda(M_Z)$, as might be expected. Our points fall mostly within a range of $\lambda(M_Z) \in (0.5, 0.8)$ with moderate $\tan\beta$. 
%There are two primary regions of interest in which we obtain alignment and large values of $\lambda(M_{GUT})\gtrsim2.0$ while satisfying the stop bounds: $\tan\beta \approx 1.5$ with $\lambda(M_Z) \approx 0.60$, and $\tan\beta \gtrsim 2.0$ with $\lambda(M_Z) \gtrsim 0.65$.

\subsection{Alignment in the doublet sector} \label{sec:align_doub}

The points obtained from running down from large values of $\lambda(M_{GUT})$, as required for a composite Higgs theory with a compositeness scale close to $M_{\rm GUT}$, fall close to the region required for exact alignment. To start with, we reduce the problem to an effective two Higgs doublet model by assuming heavy singlets and examine the mixing in the doublet sector; the suppression of the singlet mixing will be examined in the next section. To quantify how well the points fall within the alignment limit, we vary along $M_S$ and $X_t$ curves to examine the quantity 
\begin{equation}
	\cos(\beta - \alpha) = \frac{ -\mathcal{M}^2_{12} }{ \sqrt{ (m_H^2 - m_h^2)(m_H^2 - \mathcal{M}^2_{11})} }
\end{equation}
which reflects the mixing between the two doublets and reduces to $-\mathcal{M}^2_{12} / (\mathcal{M}^2_{22} - \mathcal{M}^2_{11})$ with $m_H^2 \approx \mathcal{M}^2_{22}$ and $m_h^2 \approx \mathcal{M}^2_{11}$. The $M_S$ vs. $X_t$ curve for each $(\tan\beta,\lambda(M_Z))$ point is determined by requiring that $\mathcal{M}^2_{11} = (125$ GeV)$^2$ up to the dominant two-loop terms. For low values of $\tan\beta$, the stop loop corrections tend to be smaller than the tree level values, and there is therefore little variation about the average value along each curve. As required from the choices made in the running, we use $M_A=300$ GeV in the calculation of $M_{22}^2$. Larger values of $M_A$ increase $\mathcal{M}_{22}^2$ and therefore decrease the mixing.

\begin{figure}
	\centering
	\includegraphics[width=0.75\textwidth]{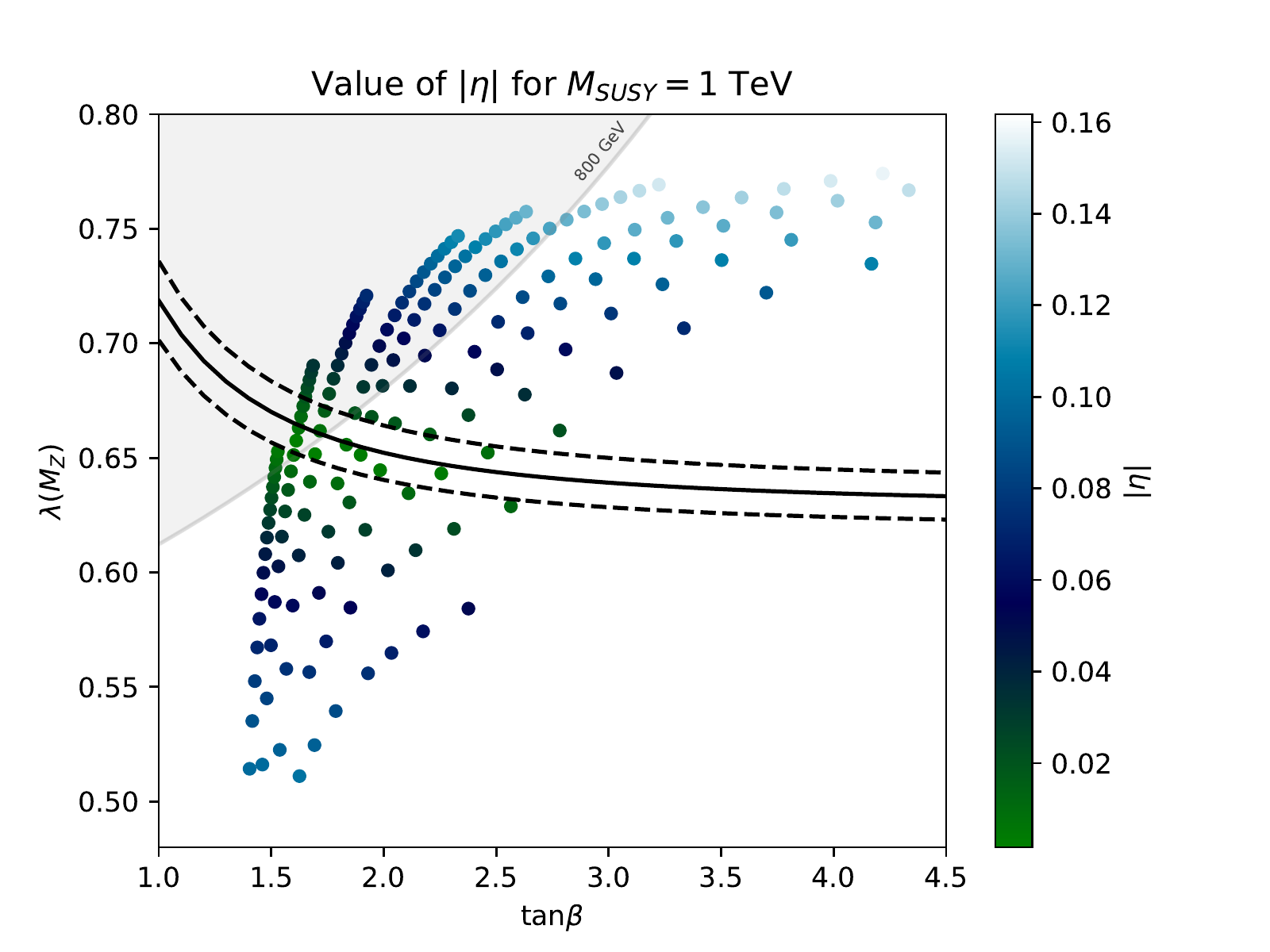}
	\caption{Values of the quantity $|\eta|$ for the points obtained from running down from $M_{GUT}$. We plot only the points which can obtain the correct Higgs mass at the 2-loop level. Points in the larger $\tan\beta$ region tend to have lower values of $\mathcal{M}_{12}^2/(\mathcal{M}_{22}^2 - \mathcal{M}^2_{11})$, but due to the larger values of $\tan\beta$ they obtain larger values of $|\eta|$ than those points at low $\tan\beta$ and $\lambda$. The shaded grey region indicates the region in which it is difficult to obtain a lighter Higgs mass of 125 GeV without tension with existing stop mass limits.}
	\label{fig:eta}
\end{figure}
  
In the effective 2HDM, the deviations of the SM-like coupling may be parametrized by~\cite{Carena:2013ooa,Carena:2014nza}
\begin{eqnarray}
g_{hb\bar{b}} & = & g_{hb\bar{b}}^{\rm SM} \left( 1 - \eta \right) \label{eq:treecoup1}\\
g_{ht\bar{t}}   & = & g_{ht\bar{t}}^{\rm SM} \left( 1 + \frac{\eta}{\tan^2\beta} \right) \label{treecoup2} \\
g_{hVV} & =  & g_{hVV}^{\rm SM} \left( 1 - \frac{\eta^2}{2 \tan^2\beta} \right) \label{eq:treecoup3}
\end{eqnarray}
where  
\begin{equation}
\eta \simeq - \tan\beta \ \frac{\mathcal{M}_{12}^2}{ \mathcal{M}_{22}^2 - \mathcal{M}_{11}^2}.
\label{eq:eta}
\end{equation}
From Eqs.~(\ref{eq:treecoup1})-(\ref{eq:treecoup3}) we see that for $\tan\beta > 1$, the tree-level bottom coupling is the one mostly affected by mixing with the non-standard states and, 
due to the relevant decay branching ratio of the SM-like Higgs to bottom quarks, it has a relevant effect on all  Higgs branching ratios. We plot the quantity $|\eta|$, which parametrizes the variation of the bottom coupling, for our weak-scale points in Fig.~\ref{fig:eta}. 

Inspection of Fig.~\ref{fig:eta} shows that the deviation of the parameter $|\eta|$ is below 0.1 for the majority of points, restricting the deviations of all couplings to values below ten percent, in agreement
with current experimental observations  \cite{ATLAS:2018doi,Sirunyan:2018koj,Aad:2019mbh} (in this work, we shall not consider the region in which the bottom Yukawa coupling acquires a wrong
sign, $\eta \simeq 2$, which can also be achieved within the NMSSM for heavy singlets~\cite{Coyle:2018ydo}).
The points on the extreme ends of the $\tan\beta$ region reach larger values of $|\eta|$, but do not exceed a deviation of 0.16. Following the same analysis with a value of $M_A=400$ GeV, we find a maximum value of $|\eta|=0.08$, which follows the expected scaling of approximately $1/M_A^2$. We therefore find that a composite Higgs model with a compositeness scale near the GUT scale may naturally lead to the alignment limit for the doublet sector at low energies. In Section~\ref{sec:FatHiggs}, we will describe a general implementation of an NMSSM Fat Higgs model with a scale $\Lambda$ of the order of $M_{\rm GUT}$.

\subsection{Alignment Condition}

As discussed above, the alignment condition in the NMSSM does not arise from a symmetry condition. To further investigate the origin of the alignment in the doublet sector, one can write the effective
two Higgs doublet potential
\bea
\label{eq:generalpotential}
V &=& m_{11}^2 \Phi_1^\dagger \Phi_1+m_{22}^2 \Phi_2^\dagger \Phi_2-m_{12}^2 (\Phi_1^\dagger \Phi_2 +{\rm h.c.}) +\frac12 \lambda_1 ( \Phi_1^\dagger \Phi_1)^2+\frac12 \lambda_2 ( \Phi_2^\dagger \Phi_2)^2 \nonumber \\
&& +\lambda_3 ( \Phi_1^\dagger \Phi_1)( \Phi_2^\dagger \Phi_2)+\lambda_4 ( \Phi_1^\dagger \Phi_2)( \Phi_2^\dagger \Phi_1) \nonumber \\
&&+ \left\{ \frac12 \lambda_5 ( \Phi_1^\dagger \Phi_2)^2 + [ \lambda_6 (\Phi^\dagger_1\Phi_1)+ \lambda_7 (\Phi^\dagger_2\Phi_2)]\Phi_1^\dagger\Phi_2 + {\rm h.c.} \right\}\ .
\eea
%For $\lambda_6 \simeq \lambda_7 \sim 0$, which is a good approximation for small values of the Higgsino mass parameter compared to the stop mass scale $\mu/M_S \ll 1$ (the dependence of the quartic couplings on the stop mass parameters is given, for instance, in Refs.~\cite{Haber:1993an,Carena:1995bx}), the condition of  alignment can be rewritten as~\cite{Carena:2013ooa}
For small values of the Higgsino mass parameter compared to the stop mass scale $\mu/M_S \ll 1$ --- the dependence of the quartic couplings on the stop mass parameters is given, for instance, in Refs.~\cite{Haber:1993an,Carena:1995bx} --- one may take $\lambda_6 \simeq \lambda_7 \sim 0$ as a good approximation. The condition of alignment can then be rewritten as~\cite{Carena:2013ooa}
\begin{eqnarray}
m_h^2 & = & \left(\lambda_1 \cos^4\beta + 2 \tilde{\lambda}_3 \sin^2\beta \cos^2\beta + \lambda_2 \sin^4\beta \right)
 \nonumber\\
m_h^2 & = & \left(\lambda_1 \cos^2\beta + \tilde{\lambda}_3 \sin^2\beta \right) v^2 ,
\end{eqnarray}
where $\tilde{\lambda}_3 = \lambda_3 + \lambda_4 + \lambda_5$.

In the literature, symmetry considerations have been invoked to relate the quartic couplings~\cite{Dev:2014yca,Dev:2017org,Benakli:2018vqz,Benakli:2018vjk,Darvishi:2019ltl}. In particular, the
condition $\lambda_1 = \lambda_2 = \tilde{\lambda}_3$ ensures alignment whenever $m_h^2 = \lambda_2 v^2$.  In the NMSSM, however, the couplings $\lambda_1$ and $\lambda_2$
differ by the sizable stop loop corrections and these conditions cannot be fulfilled. For moderate or large values of $\tan\beta \simgt 2.5$, however, the alignment conditions reduce approximately to 
$\lambda_2 \simeq \tilde{\lambda}_3$, with $m_h^2 = \lambda_2 v^2$. Taking into account that
\begin{equation}
\tilde{\lambda}_3 \simeq - \frac{g_1^2 + g_2^2}{4} + \lambda^2,
\end{equation}
one recovers the previously-obtained relation, Eq.~(\ref{lambda_align}), which in this regime of $\tan\beta$ reads
\begin{equation}
\lambda^2 \simeq \frac{M_Z^2 + m_h^2}{v^2}.
\end{equation}
Moreover,  as said above, $\lambda_2 v^2$ differs from its tree-level value $M_Z^2\simeq \lambda_1 v^2$ due to the sizable stop radiative corrections. 

The relation $\lambda_2 \simeq \tilde{\lambda}_3 \simeq m_h^2/v^2$ is therefore an emergent condition arising dynamically in the infrared limit, and it is not coming from any fundamental
symmetry.  Alignment for smaller values of $\tan\beta$ emerges in a similar way in the infrared limit. %Due to the radiative corrections induced by the top squarks, 
%the condition $\lambda_1 = \lambda_2$ is not fulfilled for any value of $\tan\beta$.  

\subsection{Alignment in the singlet sector} \label{sec:align_sing}

We must additionally examine how the mixing with the singlet Higgs might be naturally limited or suppressed due to the high-energy behavior of the theory. A similar analysis to the one performed for the doublet sector gives the exact alignment condition involving $M_{13}^2$ as
\begin{equation}
	\frac{M_A^2 s_{2\beta}^2}{4\mu^2} - \frac{\kappa s_{2\beta}}{2\lambda} = 1.
\end{equation}
For the region of $\lambda$ and $\tan\beta$ obtained by running down from the GUT scale, the value of $\sin(2\beta)$ is approximately 1. We may thusly reduce the singlet-sector alignment condition to the approximate relation
\begin{align}
	\frac{M_A^2}{4\mu^2} \approx 1,
\end{align}
where we have assumed that $\kappa/ 2\lambda$ is significantly lower than one,  as necessary to obtain a singlino state lighter than the Higgsino one, $2 \kappa/\lambda <1$, for which a natural Dark Matter
candidate may be obtained~\cite{Baum:2017enm}. 
Alignment for the singlet therefore additionally depends on the relationship between the parameters $M_A$ and $\mu$, which is not determined by the running down from $M_{GUT}$ performed above. We therefore conclude that this alignment condition cannot obviously be imposed through choices in the high-energy theory. 

We thusly examine whether one may effectively decouple the singlet due to aspects of the high-energy theory. We note that the addition of a tadpole term can effectively decouple the singlet from the doublet sector by increasing the singlet mass. In particular, the general form for $\mathcal{M}_{33}^2$ is given by~\cite{Ellwanger:2009dp}
\begin{align}
	\mathcal{M}_{33}^2 = \frac{1}{4} \lambda^2 v^2 s_{2\beta} \left( \frac{M_A^2}{2\mu^2} - \frac{\kappa}{\lambda} \right) + \frac{\kappa\mu}{\lambda} \left( A_{\kappa} + \frac{4\kappa\mu}{\lambda} \right) - \frac{\lambda}{\mu} \xi_S
\end{align}
where $\xi_S$ is the constant in a tadpole term in the Higgs potential of the form $\xi_{S} S \subset V_{H} $. A large value of $\xi_S$ can lead to large $\mathcal{M}_{33}^2$, thereby decoupling the singlet and limiting the mixing with the doublet sector. If the high-energy theory produces a singlet tadpole term in the Higgs potential, as we will examine in the next section, then the singlet mixing may be efficiently suppressed.

\section{Fat Higgs models} \label{sec:FatHiggs}

Here we focus on the possible composite nature of the Higgs, and present an example of an NMSSM Fat Higgs model~\cite{Harnik:2003rs,Chang:2004db,Delgado:2005fq} 
  which may run down to alignment at the weak scale as examined in the previous section. The primary traits we require are large values of $\lambda$ at the GUT scale and a singlet tadpole term which may decouple the singlet from mixing with the doublet sector. We therefore choose a compositeness scale of $\Lambda_H \approx M_{GUT} \approx 10^{16}$~GeV, and include a supersymmetric mass term for the two new superfields which form the singlet at low energies, thereby generating a tadpole term for $S$.

We specifically follow the construction set forth by Harnik et al. in Ref. \cite{Harnik:2003rs}, which presents an NMSSM Fat Higgs model. A new gauge symmetry $SU(2)_H$ is introduced, which becomes strong at a scale $\Lambda_H$, and six new superfields $T^{1,...6}$ are introduced which are doublets under $SU(2)_H$. $(T_1, T_2)$ also transform as a doublet under $SU(2)_L$, while $(T_3, T_4)$ and $(T_5,T_6)$ transform as singlets under $SU(2)_L$. The tree-level superpotential is given by
\begin{align}
	W =& \;\; yS'T^{1}T^{2} + yS''T^{3}T^{4} - mT^{5}T^{6} + ...
%	&+ y \begin{pmatrix} T^{1} & T^{2} \end{pmatrix} P \begin{pmatrix} T^{5} \\ T^{6} \end{pmatrix} + y \begin{pmatrix} T^{3} & T{4} \end{pmatrix} Q \begin{pmatrix} T^{5} \\ T^{6} \end{pmatrix}
\end{align}
where $S'$ and $S''$ are new singlet superfields included to ensure dynamic electroweak symmetry breaking. % while the superfields $P$ and $Q$ are matrices that combine the $T^{i}T^{j}$ pairs which form ``spectator" fields. 
Making the identifications
\begin{align} 
	S \propto T^{5}T^{6}, \;\;\;\;\;\; \begin{pmatrix} H_{u}^{+} \\ H_{u}^{0} \end{pmatrix} \propto \begin{pmatrix} T^{1}T^{3} \\ T^{2}T^{3} \end{pmatrix}, \;\;\;\;\; \begin{pmatrix} H_{d}^{0} \\ H_{d}^{-} \end{pmatrix} \propto \begin{pmatrix} T^{1}T^{4} \\ T^{2}T^{4} \end{pmatrix}
\end{align}
one obtains a dynamically-generated superpotential of 
\begin{align}
	W = \lambda S (H_u H_d - v_0^2).
\end{align}
Using Naive Dimensional Analysis~\cite{Georgi:1984iz,Georgi:1986kr,Luty:1997fk,Cohen:1997rt}, one expects that
\begin{align}
	v_0^2 &\sim \frac{m \Lambda_H}{(4\pi)^2} \\
	\lambda(\Lambda_H) &\sim 4\pi.
\end{align}

Of particular interest in our case is the very small value of $m$ required to obtain $v_0 \approx \mathcal{O}(100)$ GeV for a compositeness scale of $\Lambda_H \approx 10^{16}$ GeV; in particular, $m$ must be on the order of $10^{-1}$ eV. 

 We note that a term of the form $m T^{5}T^{6}$ may arise from the vev of a scalar superfield, in which case one would have a term of the form $g \Phi T^5 T^6$, where $g$ is a dimensionless coupling. As a scalar superfield, $\Phi$ may have the form $\Phi = \varphi + \theta \theta F$, where $\varphi$ and $F$ have some vacuum expectation values. When integrating to obtain the potential, one therefore finds an additional term linear in the Higgs singlet $S$ arising from the $F-$term. Thus, the presence of a tadpole term of the form $\xi_{F} \hat{S}$ in the superpotential may naturally give rise to a tadpole term in the potential of the form $\xi_{S} S$. 

The necessary scales can be estimated based on the values of $m$ we require due to the compositeness scale, as well as the scale of $\xi_{S}$ required to decouple the singlet from the doublet sector. We write the Higgs singlet terms with the vev of $\braket{\Phi} = \braket{\varphi} + \theta \theta \braket{F_{\phi}}$ by
\begin{align}
	g \left( \braket{\varphi} + \theta \theta \braket{F_\varphi} \right) T^{5} T^{6}
\label{eq:gPhi}	
\end{align}
where the first term generates the supersymmetric mass term $mT^{5}T^{6}$ while the second term generates the tadpole term in the potential. We estimate that $\braket{\varphi}$ and $\sqrt{|\braket{F}|}$ should both be on the order of a TeV. In order to obtain $m \sim \mathcal{O}(10^{-1})$ eV, we therefore require $g \sim \mathcal{O}(10^{-13})$. The scalar part of $\hat{S}$ then acquires a tadpole term in the potential with $\xi_{S} = \frac{\Lambda_H g \braket{F}}{4\pi}$; we require $\xi_{S}$ on the order of $10^{9}$ GeV$^3$ for decoupling, which indicates that $\Lambda_H$ is around $10^{15}$ GeV. We thus obtain a similar compositeness scale to the one that matches the NMSSM running, as described in Section \ref{sec:alignment}.

The problem now reduces to the generation of the small coupling $g$.  Such a small coupling may be explained by using a seesaw mechanism, similar to the one associated with the
Majorana neutrino mass models.  In order to propose such a model, let's follow Ref.~\cite{Harnik:2003rs} and introduce two additional $SU(2)_H$
 doublets $T^{7}$ and $T^{8}$.  We shall  assume the presence of certain flavor symmetries which
 forbid an explicit $T^{5}T^{6}$ mass term, but allow mixing between these states and the $T^7$ and $T^8$ term via  the analogue of a Giudice Masiero mechanism~\cite{Giudice:1988yz} and
 a $T^7T^8$ mass term  via the interaction with an additional superfield, $\Psi$.  Under these assumptions, the superpotential reads
 
\begin{equation}
W = \Psi T^7T^8 + m_{\rm SUSY} T^5 T^8 + m_{\rm SUSY} T^6 T^7
\end{equation}
where the $m_{\rm SUSY}$ term comes from the Giudice Masiero relation between the 
effective bilinear superfield term and the supersymmetry breaking scale.  We shall assume that 
\begin{equation}
\braket{\Psi} \simeq M + F \theta^2
\end{equation}
where $F$ is proportional to the square of the supersymmetry breaking scale, such that the superpartner masses $m_{\rm SUSY} \simeq F/M_{\rm GUT}$,  
and $M$ is of the order of the GUT scale.  Integrating out the heavy superfields $T^{7}$ and $T^{8}$, one can identify the supersymmetry conserving and breaking terms that 
appear at low energies. This can be done diagrammatically. For instance, the supersymmetry breaking tadpole term may be obtained  by considering the presence of the scalar mixing terms in the scalar potential,
\begin{equation}
V \simeq M^2 (T_7 T_7^* + T_8 T_8^*) + m_{\rm SUSY} M (T_5 T_7^* + T_6 T_8^*) + F T_7 T_8 + h.c.  ,
\end{equation}
where the first four terms arise from F terms in the superpotential, of the form $|\partial W / \partial T_7|^2$ and $|\partial W / \partial T_8|^2$, and we replace $\Psi$ by its vacuum expectation value. After integrating out the heavy fields,  the above terms lead to a supersymmetry breaking  term
\begin{equation}
V \simeq m_{\rm SUSY}^2 \frac{F}{M^2} T_5 T_6 + h.c..
\label{eq:nonSUSYtadpole}
\end{equation}
This induces a tadpole of the right size for the scalar component of $S$.

Alternatively, one  can also obtain the same result by doing a simple expansion considering the supersymmetry breaking terms like 
a perturbation of the values obtained in the supersymmetric limit.  Let's start with the supersymmetric case, with superpotential
\begin{equation}
W = M T^7T^8 + m_{\rm SUSY} T^5 T^8 + m_{\rm SUSY} T^6 T^7.
\end{equation}
Integrating out the heavy superfields, we get the effective superpotential
\begin{equation}
W = -\frac{m_{\rm SUSY}^2}{M} T_5T_6.
\end{equation}
This term, together with the supersymmetry breaking term, Eq.(\ref{eq:nonSUSYtadpole}), leads to the 
supersymmetric and non-supersymmetric tadpole contributions of the singlet $S$. 
We can then formally identify the spectator field $\Phi$ introduced in Eq.~(\ref{eq:gPhi})
with
\begin{equation}
g \braket{\Phi} \simeq - \frac{m_{\rm SUSY}^2}{\braket{\Psi}},
\end{equation}
where the above expression acquires meaning after decoupling the heavy superfields $T^7,T^8$ and performing the above mentioned expansion~\cite{Giudice:1997ni}, from which
we obtain
\begin{eqnarray}
g & \simeq & -\frac{m_{\rm SUSY}}{M}, \nonumber\\
 \braket{\Phi} & = & m_{\rm SUSY} - \frac{m_{\rm SUSY} F}{M} \theta^2 \sim m_{\rm SUSY} - m_{\rm SUSY}^2 \theta^2 .
\end{eqnarray}  
Hence, we reproduce the diagrammatic result for the supersymmetry breaking tadpole and obtain the required values of the coupling and the effective superfield 
$\Phi$ vacuum expectation values in a natural way.

While the interactions of the singlet field $S$ with the Higgs field have the required structure to obtain alignment, the
self interactions of $S$ are not determined in a clear way from our discussion above.  We shall assume that the
flavor symmetries forbid a superpotential mass term for $S$ but allow the presence of a cubic term in $S$, induced
by strong interactions at the scale $M$ and characterized
by the usual $\kappa$ term at low energies.
% which is naturally of order one in this framework.  
As shown in Appendix \ref{sec:kappa}, provided
$\kappa$ acquires moderate values there is no modification of the range of values of $\lambda$ obtained in the running. 

\subsection{Phenomenological Properties} \label{sec:Phenomenology}

The low energy limit of the model presented above is equivalent to the $\mathbb{Z}_3$ invariant NMSSM, with the addition of
tadpole terms that lift the scalar components of the singlet fields to values larger than the weak scale, implying the
suppression of the mixing of the singlet with the SM Higgs bosons. Moreover, the values of $\lambda$
ensure approximate alignment in the doublet Higgs sector.  The combination of alignment in the 
doublet Higgs sector with the decoupling of the singlet scalar fields imply that the observed Higgs boson has approximate
Standard Model-like properties, in agreement with experiments.

This model does not predict the exact value of the non-standard Higgs boson masses, but the moderate values of
$\tan\beta$ imply that the production cross section is governed by top-Yukawa induced processes. Due to the
alignment condition, which suppresses the decay into pairs of weak gauge bosons or SM-like Higgs bosons~\cite{Carena:2015moc}, 
and the absence
of light singlets, the non-standard Higgs bosons decay mostly into fermion states. Therefore,  the decay
branching ratio depends on whether the decay into pairs of top-quarks and electroweakinos is allowed.  
If top-quark decay
is dominant, searches for the heavy Higgs doublets become difficult due to interference effects with the large
top-quark production background~\cite{Dicus:1994bm}--\cite{Carena:2016npr}.  
Therefore,  the only region that is currently constrained is for low values of
$\tan\beta < 2$ and values of the heavy Higgs mass below about 350~GeV, where the top-quark decay process is
absent. The main constraint comes from the decay of the heavy Higgs bosons into $\tau$ pairs~\cite{Aaboud:2017sjh,Sirunyan:2018zut}
which, however,
can be efficiently suppressed if the electroweakinos are light~\cite{Bahl:2018zmf}.

Regarding the chargino and neutralino sectors, the model provides an acceptable Dark Matter candidate in
terms of the lightest neutralino~\cite{Ellwanger:2009dp}. Assuming this particle to be either predominantly Bino or singlino, spin 
independent direct detection bounds may be efficiently suppressed provided~\cite{Baum:2017enm}
\begin{equation}
m_{\chi} \sim \pm \mu \sin2\beta ,
\end{equation}
where the plus sign corresponds to the singlino case, while the minus sign corresponds to the Bino case. However, the suppression of the direct detection cross section in the singlino case relies on the interference between the SM-like and singlet scalar Higgs contributions, which requires a light scalar singlet. In the case of singlet decoupling, it is difficult to obtain a small direct detection cross section. However, the Bino case remains viable under direct detection limits. Moreover, values of the singlino mass close to the Bino mass and somewhat lower than the Higgsino mass $\mu$ ensure the obtention
of the proper relic density via co-annihilation of the Bino with the singlino. An acceptable relic density may therefore be obtained consistently with the
condition of avoiding the direct detection bounds in this model. Using NMSSMTools \cite{Ellwanger:2004xm} we have verified that one may indeed obtain approximate alignment with an acceptable Dark Matter candidate, for instance for $\tan\beta \simeq 2.5$ and $\lambda(M_Z) \simeq 0.69$ with values of $M_1 = 240$ GeV, $M_A \simeq 400$~GeV and $\mu = -300$ GeV, $\kappa \simeq 0.33$ and $M_S \simeq 800$~GeV (or
equivalently $m_{\tilde{t}_R} \simeq 600$~GeV and $m_{\tilde{t}_L} \simeq 1.1$~TeV). 

A further phenomenological consideration is the charged Higgs contribution to the $b \to s \gamma$ rate. Within a basic Type II 2HDM model, a light charged Higgs on the order of a few hundred GeV enhances $b \to s \gamma$ rates and therefore becomes constrained by experimental measurements \cite{Hewett:1992is,Misiak:2015xwa,Misiak:2017zan,Misiak:2017bgg}. However, within supersymmetric theories these flavor rates also depend strongly on the contributions from other supersymmetric particles; these include charginos and stops, which can exactly cancel the SM contributions to the $b \to s \gamma$ transition in the limit of exact supersymmetry \cite{Barbieri:1993av,Degrassi:2000qf,Carena:2000uj,Buras:2002vd}. Furthermore, there are contributions arising from possible flavor violation in the scalar fermion sector; these can be large corrections arising from gluino-squark loops. This can occur when there is a misalignment between the bases in which the quark and squark mass matrices are diagonalized \cite{Gabbiani:1996hi}. In light of this, we do not further consider flavor constraints; however, we have confirmed using NMSSMTools that the models described above can be in agreement with flavor constraints up to the SUSY contributions included in NMSSMTools.

\section{Unification of $h_b$ and $h_{\tau}$}
\label{sec:Unification}

\begin{figure}[H]
	\centering
	\includegraphics[width=0.75\textwidth]{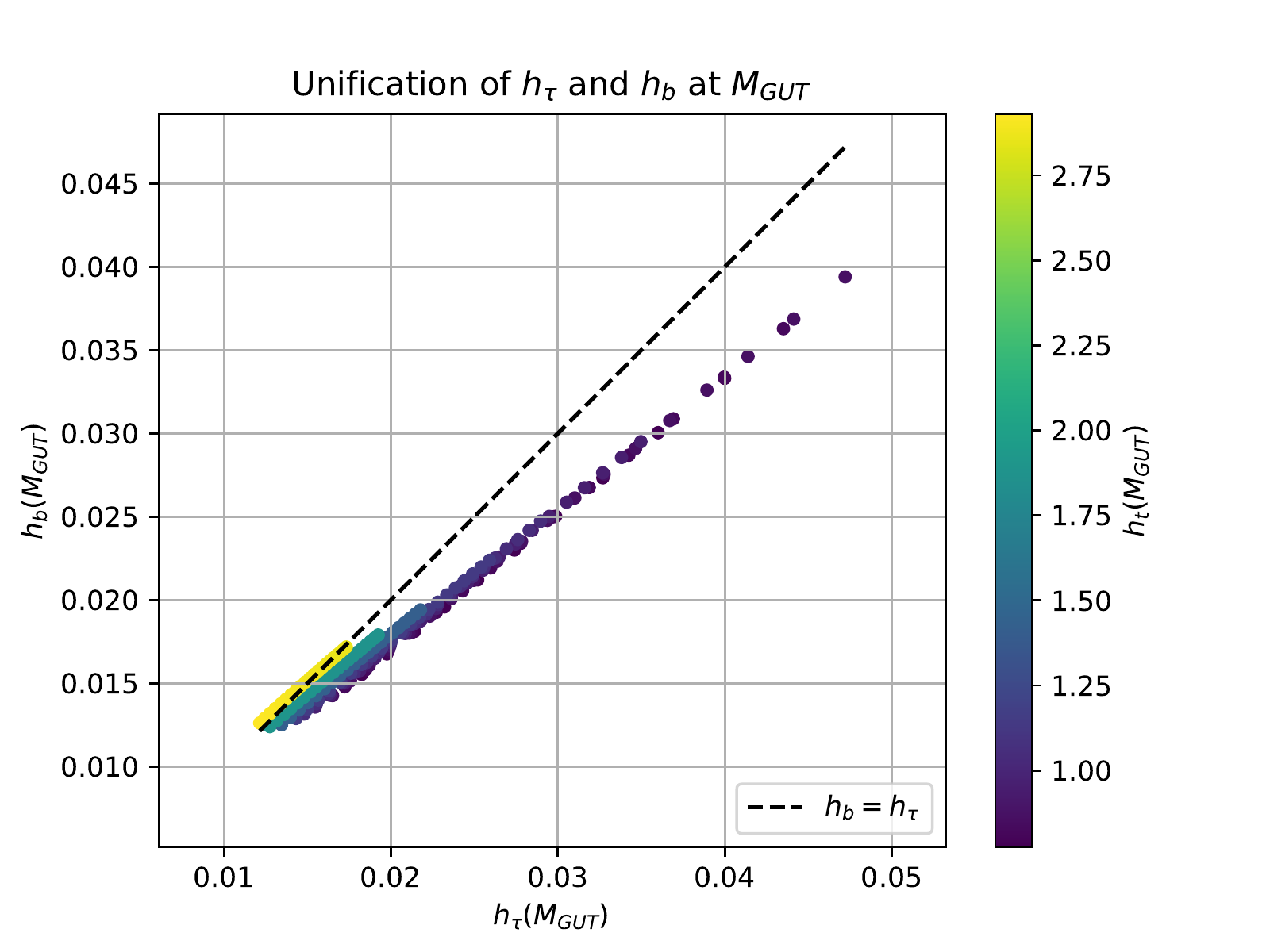}
	\caption{Plot of the values of $h_{\tau}(M_{GUT})$ and $h_b(M_{GUT})$ obtained from running the weak-scale points shown in Fig.~\ref{contour_plots} up to the GUT scale. The color bar indicates the value of $h_t(M_{GUT})$, for which larger values push the values of $h_{\tau}$ and $h_b$ closer to unification at the GUT scale.}
	\label{btau_unification}
\end{figure}

Although it is not directly related to the alignment in the Higgs sector, 
another intriguing aspect of the running of the RG evolution from the alignment limit is the unification of $h_b$ and $h_{\tau}$ at the GUT scale. Fig.~\ref{btau_unification} shows the values of $h_{b}(M_{GUT})$ and $h_{\tau}(M_{GUT})$ obtained by running the weak-scale points in Fig.~\ref{contour_plots} upward to the GUT scale. %The differently-colored lines of points are differentiated by the value of $h_{t}$ at the GUT scale, while each point along a line represents a different value of $\lambda$.
As expected from previous work~\cite{Arason:1991hu}--\cite{Kolda:1994ab},
for such large values of $h_t(M_{GUT})$ the values approach the $h_b(M_{GUT}) = h_{\tau}(M_{GUT})$ line as $h_t$ increases.  The values of $h_t$ approach an infrared fixed point~\cite{Bardeen:1993rv}, which is 
also a feature of top condensate models~\cite{Bardeen:1989ds,Clark:1989tq,Carena:1991ky}, which is a different realization of compositeness in the Higgs doublet sector. 

The unification of the bottom and tau Yukawa couplings suggests that the bottom-quark and $\tau$-lepton share  the same representations of the high-energy theory, as would happen in an  effective $SU(5)$ theory at the GUT scale.  However, GUT scenarios tend to encounter a number of phenomenological issues (see, for example, Refs.~\cite{Komine:2001rm}--\cite{Baer:2012by}), and an examination of how one may build a successful grand unification theory with the NMSSM as the low-energy theory, along with composite Higgs bosons, is beyond the scope of this paper.

\section{Conclusion} \label{sec:Conclusion}

The condition of alignment in the Higgs sector allows for the possibility of obtaining a relatively light Higgs spectrum without 
being in conflict with the LHC Higgs precision measurements.  Quite generally, alignment is not dictated by any symmetry
consideration and, barring the possibility of being an accidental condition, requires a dynamical explanation. In this article
we concentrated on the NMSSM, in which the alignment condition is associated with a narrow range of values of the 
superpotential coupling $\lambda$, which governs the interactions of the singlet to doublet Higgs states.  
For low values of $\tan\beta \simlt 3$, this range of values of $\lambda$ leads to the observed value of the Higgs mass without
requiring a very large stop spectrum. Moreover, as shown in this article,  the renormalization group evolution of the coupling 
$\lambda$ shows that for low energy values which lead to alignment in the Higgs sector,  $\lambda$ tends to become strong at scales of 
the order of the GUT scale.  Furthermore, the top Yukawa coupling also tends to large values at similar large energy scales. 

In this article we interpret the large values of $\lambda$ at the GUT scale as a signal of compositeness of the Higgs states. 
Following this idea, we construct a Fat Higgs Model with a compositeness scale that is close to the GUT scale, which leads 
to the desired Higgs spectrum and allows for the  presence of a tadpole contribution that leads to the natural decoupling of the 
singlet scalar states in the low energy theory.  This implies that the alignment in the doublet sector, governed by $\lambda$,
ensures the SM-like properties of the lightest Higgs, as required by the LHC precision measurements. 

In addition to obtaining a Higgs sector which is consistent with current experimental constraints, the model also includes
a Dark Matter candidate, which is mostly Bino-like and obtains the correct relic density through coannihilation with light singlinos.  
Moreover, for values  of the Dark Matter mass close to $- \mu \sin 2\beta$, direct detection constraints can be avoided in the 
Bino case. All these conditions may be simultaneously satisfied within these models.   

Finally, we stress that the relatively strong values of the top Yukawa coupling lead to the unification of the bottom and
tau Yukawa couplings at the GUT scale.  This suggests the possible embedding of this theory within a GUT scenario,
like $SU(5)$, in which the bottom-quark and tau-lepton share  the same multiplets.  We reserve for future work the 
construction of such a theory.

\section{Acknowledgements}

We thank Timothy Tait, Antonio Delgado, Nausheen Shah, and Roni Harnik for useful discussions. Work at ANL is supported in part by the U.S. Department of Energy under Contract No. DE-AC02-06CH11357. The work of C.W. and N.C. at EFI is supported by the U.S. Department of Energy under Contract No. DE-FG02-13ER41958. Work by N.C. at FNAL is supported by the U.S. Department of Energy, Office of Science, Office of Workforce Development for Teachers and Scientists, Office of Science Graduate Student Research (SCGSR) program. The SCGSR program is administered by the Oak Ridge Institute for Science and Education (ORISE) for the DOE. ORISE is managed by ORAU under contract number DE-SC0014664.

\newpage

\appendix

\section{RGE equations}
\label{sec:RGE}

We list here the RGE equations used for the running of the couplings presented in Section \ref{sec:alignment}. 

\subsection{SM and 2HDM}

The equations used for the Standard Model and 2HDM running are given by~\cite{Branco:2011iw}

\begin{align}
	\frac{d\alpha_3}{dt} &= 7 \frac{\alpha_3^2}{4\pi} \nonumber \\
	\frac{d\alpha_2}{dt} &= \beta_2 \frac{\alpha_2^2}{4\pi} \nonumber \\
	\frac{d\alpha_1}{dt} &= -\beta_1 \frac{\alpha_1^2}{4\pi} \nonumber \\
	\frac{dY_t}{dt} &= Y_t \left( 8 \tilde{\alpha}_3 + \frac{9}{4} \tilde{\alpha}_2 + \frac{17}{12}\tilde{\alpha}_1 - \frac{9}{2}Y_t - \frac{\alpha_b}{2}Y_b  - \alpha_{\tau} Y_{\tau} \right)  \\
	\frac{dY_b}{dt} &= Y_b \left( 8 \tilde{\alpha}_3 + \frac{9}{4} \tilde{\alpha}_2 + \frac{5}{12} \tilde{\alpha}_1 - \frac{9}{2}Y_b - \frac{\alpha_t}{2}Y_t  - Y_\tau \right) \nonumber \\
	\frac{dY_{\tau}}{dt} &= Y_{\tau} \left( \frac{9}{4} \tilde{\alpha}_2 + \frac{15}{4} \tilde{\alpha}_1 - \frac{5}{2} Y_{\tau} - 3 Y_b - \alpha'_t Y_t \right) \nonumber
\end{align}
where $\alpha_i = g_i^2/4\pi$, $\tilde{\alpha}_i = \alpha_i/4\pi$, $Y_{t,b} = h_{t,b}^2/16\pi^2$, and $t=\log(M_{GUT}^2/\mu^2)$. The parameters $(\beta_2, \beta_1, \alpha_b, \alpha_t, \alpha'_t,\alpha_\tau)$ are equal to $(3, 7, 1, 1, 0,0)$ for the 2HDM and $(19/6, 41/6, 3, 3, 3,1)$ for the SM running.

\subsection{NMSSM}

The 2-loop RGE~\cite{Ellwanger:2009dp} used for the NMSSM running are listed below; we employ the SM normalization of $g_1$. The running parameter $t$ is defined here as $t = \ln \left( Q^2 / M_Z^2 \right)$.

\begin{align}
	16 \pi^2 \frac{d g_1^2}{dt} &= 11 g_1^4 + \frac{g_1^4}{16\pi^2} \left( \frac{199}{9} g_1^2 + 9 g_2^2 + \frac{88}{3} g_3^2 - \frac{26}{3} h_t^2 - \frac{14}{3} h_b^2 - 6 h_{\tau}^2 - 2\lambda^2 \right) \nonumber \\
	16\pi^2 \frac{d g_2^2}{dt} &= g_2^4 + \frac{g_2^4}{16\pi^2} \left( 3 g_1^2 + 25 g_2^2 + 24 g_3^2 - 6 h_t^2 - 6 h_b^2 - 2 h_{\tau}^2 - 2\lambda^2 \right) \nonumber \\
	16\pi^2 \frac{d g_3^2}{dt} &= -3 g_3^4 + \frac{g_3^4}{16\pi^2} \left( \frac{11}{3} g_1^2 + 9 g_2^2 + 14 g_3^2 - 4 h_t^2 - 4 h_b^2 \right) \nonumber \\
	16\pi^2 \frac{d h_t^2}{dt} &= h_t^2 \left( 6 h_t^2 + h_b^2 + \lambda^2 - \frac{13}{9} g_1^2 - 3 g_2^2 - \frac{16}{3} g_3^2 \right) \nonumber \\
	&+ \frac{h_t^2}{16\pi^2} \Bigg( -22 h_t^4 - 5 h_b^4 - 3\lambda^4 - 5 h_t^2 h_b^2 - 3 h_t^2 \lambda^2 - h_b^2 h_{\tau}^2 - 4 h_b^2 \lambda^2  \nonumber \\
	& - h_{\tau}^2 \lambda^2 - 2 \lambda^2 \kappa^2 + 2 g_1^2 h_t^2 + \frac{2}{3} g_1^2 h_b^2 + 6 g_2^2 h_t^2 + 16 g_3^2 h_t^2 \nonumber \\
	&+ \frac{2743}{162} g_1^4 + \frac{15}{2} g_2^4 - \frac{16}{9} g_3^4 + \frac{5}{3} g_1^2 g_2^2 + \frac{136}{27} g_1^2 g_3^2 + 8 g_2^2 g_3^2 \Bigg) \nonumber \\
	16\pi^2 \frac{d h_b^2}{dt} &= h_b^2 \left( 6 h_b^2 + h_t^2 + h_{\tau}^2 + \lambda^2 - \frac{7}{9} g_1^2 - 3 g_2^2 - \frac{16}{3} g_3^2 \right) \nonumber \\
	&+ \frac{h_b^2}{16\pi^2} \Bigg( -22 h_b^4 - 5 h_t^4 - 3 h_{\tau}^4 - 3 \lambda^4 - 5 h_b^2 h_t^2 - 3 h_b^2 h_{\tau}^2 - 3 h_b^2 \lambda^2 \nonumber \\
	&- 4 h_t^2 \lambda^2 - 2 \lambda^2 \kappa^2 + \frac{2}{3} g_1^2 h_b^2 + \frac{4}{3} g_1^2 h_t^2 + 2 g_1^2 h_{\tau}^2 + 6 g_2^2 h_b^2 + 16 g_3^2 h_b^2 \nonumber \\
	&+ \frac{1435}{162} g_1^4 + \frac{15}{2} g_2^4 - \frac{16}{9} g_3^4 + \frac{5}{3} g_1^2 g_2^2 + \frac{40}{27} g_1^2 g_3^2 + 8 g_2^2 g_3^2 \Bigg) \nonumber \\
	16 \pi^2 \frac{d h_{\tau}^2}{dt} &= h_{\tau}^2 \left( 4 h_{\tau}^2 + 3 h_b^2 + \lambda^2 - 3 g_1^2 - 3 g_2^2 \right) \nonumber \\
	&+ \frac{h_{\tau}^2}{16\pi^2} \Bigg( -10 h_{\tau}^4 - 9 h_b^4 - 3 \lambda^4 - 9 h_{\tau}^2 h_b^2 - 3 h_{\tau}^2 \lambda^2 - 3 h_t^2 h_b^2 - 3 h_t^2 \lambda^2 \nonumber \\
	&- 2 \lambda^2 \kappa^2 + 2 g_1^2 h_{\tau}^2 - \frac{2}{3} g_1^2 h_b^2 + 6 g_2^2 h_{\tau}^2 + 16 g_3^2 h_b^2 + \frac{75}{2} g_1^4 + \frac{15}{2} g_2^4 + 3 g_1^2 g_2^2 \Bigg) \nonumber \\
	16 \pi^2 \frac{d \lambda^2}{dt} &= \lambda^2 \left( 3 h_t^2 + 3 h_b^2 + h_{\tau}^2 + 4 \lambda^2 + 2 \kappa^2 - g_1^2 - 3 g_2^2 \right) \nonumber \\
	&+ \frac{\lambda^2}{16\pi^2} \Bigg( -10 \lambda^4 - 9 h_t^4 - 9 h_b^4 - 3 h_{\tau}^4 - 8 \kappa^4 - 9 \lambda^2 h_t^2 - 9 \lambda^2 h_b^2 \nonumber \\
	&- 3 \lambda^2 h_{\tau}^2 - 12 \lambda^2 \kappa^2 - 6 h_t^2 h_b^2 + 2 g_1^2 \lambda^2 + \frac{4}{3} g_1^2 h_t^2 - \frac{2}{3} g_1^2 h_b^2 + 2 g_1^2 h_{\tau}^2 \nonumber \\
	&+ 6 g_2^2 \lambda^2 + 16 g_3^2 h_t^2 + 16 g_3^2 h_b^2 + \frac{23}{2} g_1^4 + \frac{15}{2} g_2^4 + 3 g_1^2 g_2^2 \Bigg) \nonumber \\
	16 \pi^2 \frac{d \kappa^2}{dt} &= \kappa^2 \left( 6\lambda^2 + 6\kappa^2 \right) + \frac{\kappa^2}{16\pi^2} \Big( -24 \kappa^4 - 12 \lambda^4 - 24 \kappa^2 \lambda^2 \nonumber \\
	&- 18 h_t^2 \lambda^2 - 18 h_b^2 \lambda^2 - 6 h_{\tau}^2 \lambda^2 + 6 g_1^2 \lambda^2 + 18 g_2^2 \lambda^2 \Big) \nonumber
\end{align}

%\section{Additional plots for $M_{SUSY}=5$ TeV and 10 TeV}

%Here we present the plots corresponding to Fig. \ref{M22_align} and Fig. \ref{btau_unification} for SUSY scales of 5 TeV and 10 TeV.

%\begin{figure}[H]
%	\centering
%	\includegraphics[width=0.48\textwidth]{../Plots/Contour_plots/alignment_5k_2HDM}
%	\includegraphics[width=0.48\textwidth]{../Plots/Contour_plots/alignment_10k_2HDM}
%	\caption{Plots showing the alignment of points obtained for $M_{SUSY}$=5 TeV and 10 TeV.}
%\end{figure}

%\begin{figure}[H]
%	\centering
%	\includegraphics[width=0.48\textwidth]{../Plots/Coupling_unification_plots/htauhb_5k_2HDM}
%	\includegraphics[width=0.48\textwidth]{../Plots/Coupling_unification_plots/htauhb_10k_2HDM}
%	\caption{Plots showing the unification of $h_b$ and $h_{\tau}$ for $M_{SUSY}$=5 TeV and 10 TeV.}
%\end{figure}

\section{Results for non-zero $\kappa$} \label{sec:kappa}

In the main analysis of this paper, we set $\kappa=0$; in this section we present results for different values of $\kappa$ at the GUT scale to quantify the effect of a non-zero value of $\kappa$ on the results of the running. Here we plot the results of running downward for $\kappa(M_{GUT})=1,2$ with $M_{SUSY}=1$ TeV.

\begin{figure}[H]
	\centering
	\includegraphics[width=0.48\textwidth]{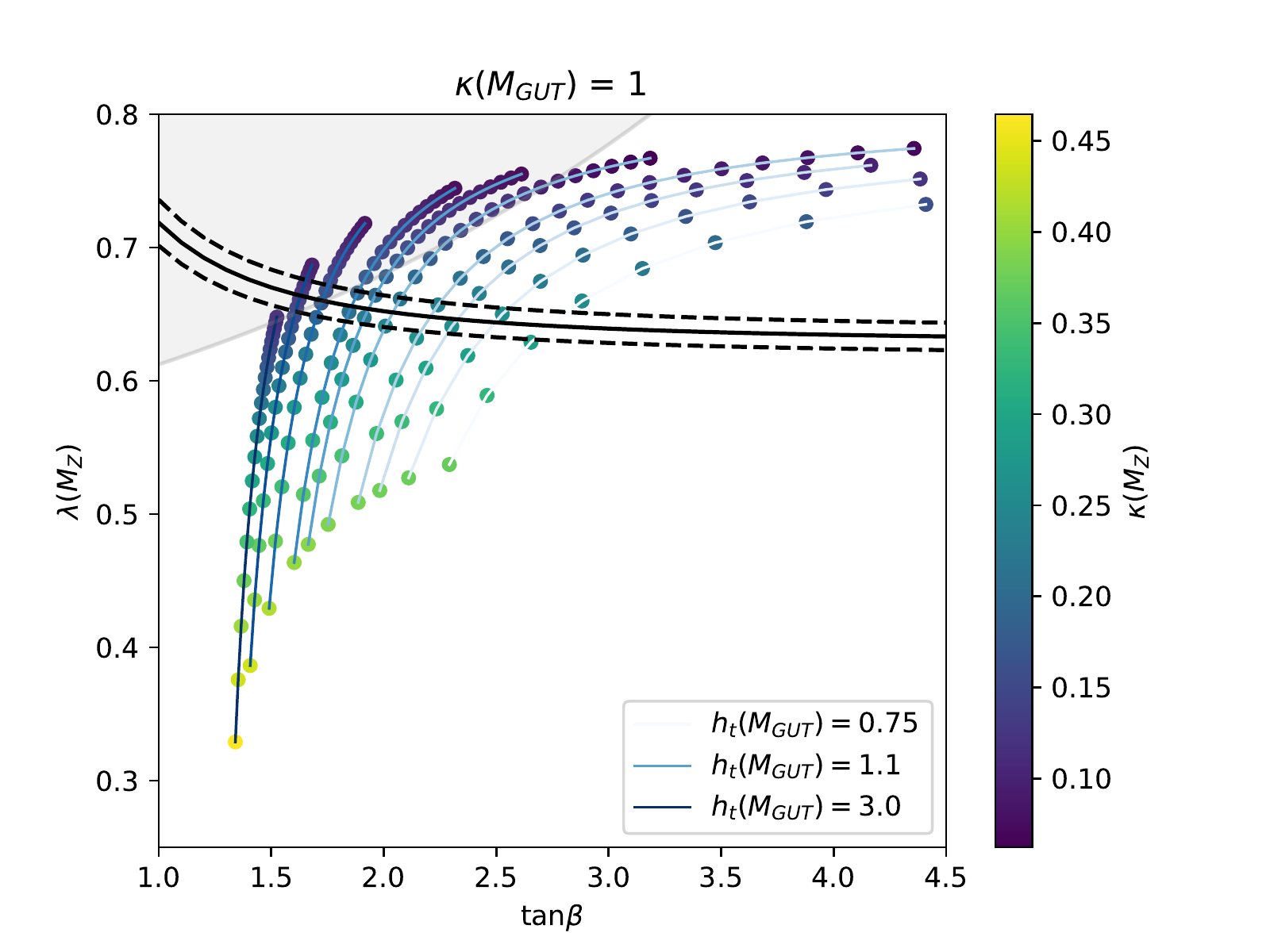} \;
	\includegraphics[width=0.48\textwidth]{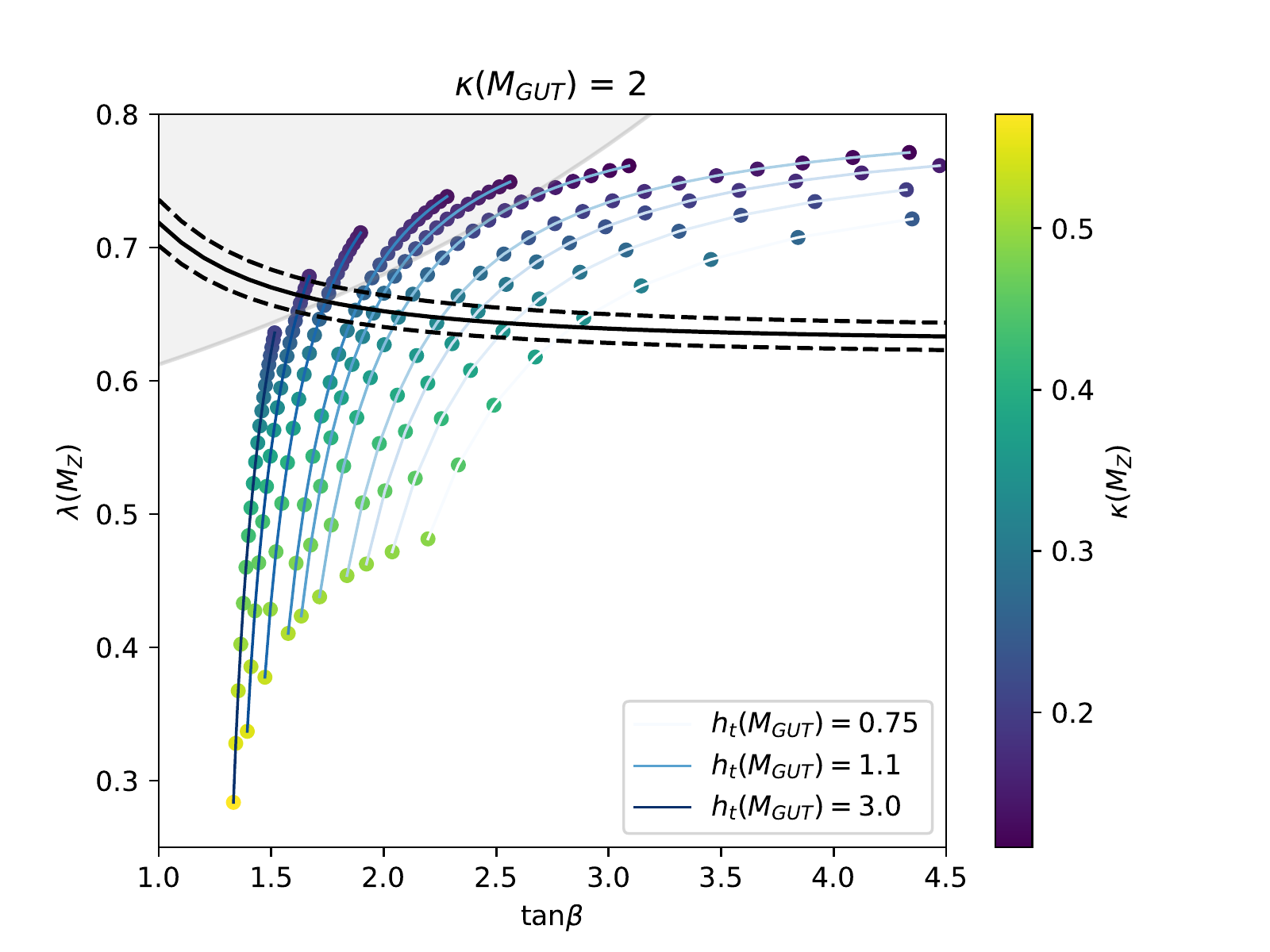}
	\caption{Plots showing the points obtained from running down from the GUT scale with different values of $\kappa(M_{GUT})$. All other parameter choices, including the ranges of $\lambda(M_{GUT})$ and $h_t(M_{GUT})$, are the same as for Fig.~\ref{contour_plots}.}
	\label{fig:kap1}
\end{figure}

The large values of $\kappa(M_{GUT})$ do not significantly affect the weak-scale parameter values, which remain near the alignment region. The primary effect of increased $\kappa$ is a lower value of $\lambda(M_Z)$, which tends to be reduced by up to about 0.05 relative to the $\kappa=0$ case. 
%We plot the results for $\kappa = 0, 0.5, 1, 4$ together in a single plot in Fig.~\ref{fig:kap2} to directly compare the variation in the values of $\lambda(m_Z)$ and $\tan\beta$ obtained with different values of $\kappa$.
Based on the low variation in the results with large $\kappa$, we conclude that setting $\kappa=0$ provides a representative analysis.


\newpage

\begin{thebibliography}{99}

%\cite{Aad:2012tfa}
\bibitem{Aad:2012tfa} 
  G.~Aad {\it et al.} [ATLAS Collaboration],
  %``Observation of a new particle in the search for the Standard Model Higgs boson with the ATLAS detector at the LHC,''
  Phys.\ Lett.\ B {\bf 716}, 1 (2012)
  doi:10.1016/j.physletb.2012.08.020
  [arXiv:1207.7214 [hep-ex]].
  %%CITATION = doi:10.1016/j.physletb.2012.08.020;%%
  %9615 citations counted in INSPIRE as of 12 Aug 2019
  
  %\cite{Aad:2012tfa}
  %\cite{Chatrchyan:2012xdj}
\bibitem{Chatrchyan:2012xdj} 
  S.~Chatrchyan {\it et al.} [CMS Collaboration],
  %``Observation of a New Boson at a Mass of 125 GeV with the CMS Experiment at the LHC,''
  Phys.\ Lett.\ B {\bf 716}, 30 (2012)
  doi:10.1016/j.physletb.2012.08.021
  [arXiv:1207.7235 [hep-ex]].
  %%CITATION = doi:10.1016/j.physletb.2012.08.021;%%
  %9402 citations counted in INSPIRE as of 12 Aug 2019
  
  %\cite{Aad:2015zhl}
\bibitem{Aad:2015zhl} 
  G.~Aad {\it et al.} [ATLAS and CMS Collaborations],
  %``Combined Measurement of the Higgs Boson Mass in $pp$ Collisions at $\sqrt{s}=7$ and 8 TeV with the ATLAS and CMS Experiments,''
  Phys.\ Rev.\ Lett.\  {\bf 114}, 191803 (2015)
  doi:10.1103/PhysRevLett.114.191803
  [arXiv:1503.07589 [hep-ex]].
  %%CITATION = doi:10.1103/PhysRevLett.114.191803;%%
  %1395 citations counted in INSPIRE as of 12 Aug 2019
  
 
  %\cite{Gunion:2002zf}
\bibitem{Gunion:2002zf} 
  J.~F.~Gunion and H.~E.~Haber,
  %``The CP conserving two Higgs doublet model: The Approach to the decoupling limit,''
  Phys.\ Rev.\ D {\bf 67}, 075019 (2003)
  doi:10.1103/PhysRevD.67.075019
  [hep-ph/0207010].
  %%CITATION = doi:10.1103/PhysRevD.67.075019;%%
  %656 citations counted in INSPIRE as of 12 Aug 2019
  
 

  
  %\cite{Delgado:2013zfa}
\bibitem{Delgado:2013zfa} 
  A.~Delgado, G.~Nardini and M.~Quiros,
  %``A Light Supersymmetric Higgs Sector Hidden by a Standard Model-like Higgs,''
  JHEP {\bf 1307}, 054 (2013)
  doi:10.1007/JHEP07(2013)054
  [arXiv:1303.0800 [hep-ph]].
  %%CITATION = doi:10.1007/JHEP07(2013)054;%%
  %53 citations counted in INSPIRE as of 12 Aug 2019
  
  %\cite{Craig:2013hca}
\bibitem{Craig:2013hca} 
  N.~Craig, J.~Galloway and S.~Thomas,
  %``Searching for Signs of the Second Higgs Doublet,''
  arXiv:1305.2424 [hep-ph].
  %%CITATION = ARXIV:1305.2424;%%
  %233 citations counted in INSPIRE as of 12 Aug 2019
  
  %\cite{Carena:2013ooa}
\bibitem{Carena:2013ooa} 
  M.~Carena, I.~Low, N.~R.~Shah and C.~E.~M.~Wagner,
  %``Impersonating the Standard Model Higgs Boson: Alignment without Decoupling,''
  JHEP {\bf 1404}, 015 (2014)
  doi:10.1007/JHEP04(2014)015
  [arXiv:1310.2248 [hep-ph]].
  %%CITATION = doi:10.1007/JHEP04(2014)015;%%
  %166 citations counted in INSPIRE as of 08 Aug 2019
  
  
  %\cite{Carena:2014nza}
\bibitem{Carena:2014nza} 
  M.~Carena, H.~E.~Haber, I.~Low, N.~R.~Shah and C.~E.~M.~Wagner,
  %``Complementarity between Nonstandard Higgs Boson Searches and Precision Higgs Boson Measurements in the MSSM,''
  Phys.\ Rev.\ D {\bf 91}, no. 3, 035003 (2015)
  doi:10.1103/PhysRevD.91.035003
  [arXiv:1410.4969 [hep-ph]].
  %%CITATION = doi:10.1103/PhysRevD.91.035003;%%
  %71 citations counted in INSPIRE as of 08 Aug 2019
  
  
  %\cite{Carena:2015moc}
\bibitem{Carena:2015moc} 
  M.~Carena, H.~E.~Haber, I.~Low, N.~R.~Shah and C.~E.~M.~Wagner,
  %``Alignment limit of the NMSSM Higgs sector,''
  Phys.\ Rev.\ D {\bf 93}, no. 3, 035013 (2016)
  doi:10.1103/PhysRevD.93.035013
  [arXiv:1510.09137 [hep-ph]].
  %%CITATION = doi:10.1103/PhysRevD.93.035013;%%
  %31 citations counted in INSPIRE as of 03 Dec 2018
  
  %\cite{Bernon:2015qea}
\bibitem{Bernon:2015qea} 
  J.~Bernon, J.~F.~Gunion, H.~E.~Haber, Y.~Jiang and S.~Kraml,
  %``Scrutinizing the alignment limit in two-Higgs-doublet models: m$_h$=125  GeV,''
  Phys.\ Rev.\ D {\bf 92}, no. 7, 075004 (2015)
  doi:10.1103/PhysRevD.92.075004
  [arXiv:1507.00933 [hep-ph]].
  %%CITATION = doi:10.1103/PhysRevD.92.075004;%%
  %120 citations counted in INSPIRE as of 26 Nov 2019
  
    %\cite{Haber:2017erd}
\bibitem{Haber:2017erd} 
  H.~E.~Haber, S.~Heinemeyer and T.~Stefaniak,
  %``The Impact of Two-Loop Effects on the Scenario of MSSM Higgs Alignment without Decoupling,''
  Eur.\ Phys.\ J.\ C {\bf 77}, no. 11, 742 (2017)
  doi:10.1140/epjc/s10052-017-5243-5
  [arXiv:1708.04416 [hep-ph]].
  %%CITATION = doi:10.1140/epjc/s10052-017-5243-5;%%
  %13 citations counted in INSPIRE as of 26 Nov 2019

  
 %\cite{Dev:2014yca}
\bibitem{Dev:2014yca}  
  P.~S.~B.~Dev and A.~Pilaftsis,
  %``Maximally Symmetric Two Higgs Doublet Model with Natural Standard Model Alignment,''
  JHEP {\bf 1412}, 024 (2014)
  [arXiv:1408.3405 [hep-ph]];
  %%CITATION = ARXIV:1408.3405;%%
  %19 citations counted in INSPIRE as of 24 ao?t 2015
 %\cite{Dev:2014yca}
 %\cite{Dev:2017org}
\bibitem{Dev:2017org} 
  P.~S.~Bhupal Dev and A.~Pilaftsis,
  %``Natural Alignment in the Two Higgs Doublet Model,''
  J.\ Phys.\ Conf.\ Ser.\  {\bf 873}, no. 1, 012008 (2017)
  doi:10.1088/1742-6596/873/1/012008
  [arXiv:1703.05730 [hep-ph]].
  %%CITATION = doi:10.1088/1742-6596/873/1/012008;%%
  %4 citations counted in INSPIRE as of 26 Nov 2019
   %\cite{Dev:2014yca}
 %\cite{Dev:2017org}
 %\cite{Benakli:2018vqz}
\bibitem{Benakli:2018vqz} 
  K.~Benakli, M.~D.~Goodsell and S.~L.~Williamson,
  %``Higgs alignment from extended supersymmetry,''
  Eur.\ Phys.\ J.\ C {\bf 78}, no. 8, 658 (2018)
  doi:10.1140/epjc/s10052-018-6125-1
  [arXiv:1801.08849 [hep-ph]].
  %%CITATION = doi:10.1140/epjc/s10052-018-6125-1;%%
  %12 citations counted in INSPIRE as of 26 Nov 2019  %\cite{Benakli:2018vjk}
  %\cite{Dev:2014yca}
 %\cite{Dev:2017org}
 %\cite{Benakli:2018vqz}
%\cite{Benakli:2018vjk}
\bibitem{Benakli:2018vjk} 
  K.~Benakli, Y.~Chen and G.~Lafforgue-Marmet,
  %``R-symmetry for Higgs alignment without decoupling,''
  Eur.\ Phys.\ J.\ C {\bf 79}, no. 2, 172 (2019)
  doi:10.1140/epjc/s10052-019-6676-9
%\cite{Dev:2014yca}
 %\cite{Dev:2017org}
 %\cite{Benakli:2018vqz}
%\cite{Benakli:2018vjk}
  %\cite{Darvishi:2019ltl}
\bibitem{Darvishi:2019ltl} 
  N.~Darvishi and A.~Pilaftsis,
  %``Quartic Coupling Unification in the Maximally Symmetric 2HDM,''
  Phys.\ Rev.\ D {\bf 99}, no. 11, 115014 (2019)
  doi:10.1103/PhysRevD.99.115014
  [arXiv:1904.06723 [hep-ph]].
  %%CITATION = doi:10.1103/PhysRevD.99.115014;%%
  %3 citations counted in INSPIRE as of 13 Dec 2019
  [arXiv:1811.08435 [hep-ph]].
  %%CITATION = doi:10.1140/epjc/s10052-019-6676-9;%%
  %4 citations counted in INSPIRE as of 26 Nov 2019  
  
  
  %\cite{higgsbasis,Branco:1999fs}
 \bibitem{higgsbasis}
    %\cite{Georgi:1978ri}
%\bibitem{Georgi:1978ri} 
  H.~Georgi and D.~V.~Nanopoulos,
  %``Suppression of Flavor Changing Effects From Neutral Spinless Meson Exchange in Gauge Theories,''
  Phys.\ Lett.\ B {\bf 82}, 95 (1979);
  %%CITATION = PHLTA,B82,95;%%
  %83 citations counted in INSPIRE as of 15 Oct 2014
 %\cite{Donoghue:1978cj}
%\bibitem{Donoghue:1978cj} 
  J.~F.~Donoghue and L.~F.~Li,
  %``Properties of Charged Higgs Bosons,''
  Phys.\ Rev.\ D {\bf 19}, 945 (1979);
  %%CITATION = PHRVA,D19,945;%%
  %177 citations counted in INSPIRE as of 15 Oct 2014
  %\cite{Lavoura:1994fv}
%\bibitem{Lavoura:1994fv} 
  L.~Lavoura and J.~P.~Silva,
  %``Fundamental CP violating quantities in a SU(2) x U(1) model with many Higgs doublets,''
  Phys.\ Rev.\ D {\bf 50}, 4619 (1994)
  [hep-ph/9404276];
  %%CITATION = HEP-PH/9404276;%%
  %72 citations counted in INSPIRE as of 15 Oct 2014
   %\cite{Lavoura:1994yu}
%\bibitem{Lavoura:1994yu} 
  L.~Lavoura,
  %``Signatures of discrete symmetries in the scalar sector,''
  Phys.\ Rev.\ D {\bf 50}, 7089 (1994)
  [hep-ph/9405307];
  %%CITATION = HEP-PH/9405307;%%
  %18 citations counted in INSPIRE as of 15 Oct 2014
  %\cite{Botella:1994cs}
%\bibitem{Botella:1994cs} 
  F.~J.~Botella and J.~P.~Silva,
  %``Jarlskog - like invariants for theories with scalars and fermions,''
  Phys.\ Rev.\ D {\bf 51}, 3870 (1995)
  [hep-ph/9411288].
  %%CITATION = HEP-PH/9411288;%%
  %73 citations counted in INSPIRE as of 15 Oct 2014
  
  %\cite{Branco:1999fs}
\bibitem{Branco:1999fs} 
  See Chapter 22 of G.~C.~Branco, L.~Lavoura and J.~P.~Silva,
  \textit{CP Violation} (Oxford University Press, Oxford, UK, 1999).
 % Int.\ Ser.\ Monogr.\ Phys.\  {\bf 103}, 1 (1999).
  %%CITATION = IMPHA,103,1;%%
  %89 citations counted in INSPIRE as of 15 Oct 2014
  
  %\cite{Ellwanger:2009dp}
\bibitem{Ellwanger:2009dp} 
  U.~Ellwanger, C.~Hugonie and A.~M.~Teixeira,
  %``The Next-to-Minimal Supersymmetric Standard Model,''
  Phys.\ Rept.\  {\bf 496}, 1 (2010)
  doi:10.1016/j.physrep.2010.07.001
  [arXiv:0910.1785 [hep-ph]].
  %%CITATION = doi:10.1016/j.physrep.2010.07.001;%%
  %860 citations counted in INSPIRE as of 03 Dec 2018  
  
   
 \bibitem{mssmhiggsradcorr}
 %\cite{Haber:1990aw}
%\bibitem{Haber:1990aw} 
  H.~E.~Haber and R.~Hempfling,
  %``Can the mass of the lightest Higgs boson of the minimal supersymmetric model be larger than m(Z)?,''
  Phys.\ Rev.\ Lett.\  {\bf 66}, 1815 (1991);
  %%CITATION = PRLTA,66,1815;%%
  %1224 citations counted in INSPIRE as of 15 Oct 2014
  %\cite{Okada:1990vk}
%\bibitem{Okada:1990vk} 
  Y.~Okada, M.~Yamaguchi and T.~Yanagida,
  %``Upper bound of the lightest Higgs boson mass in the minimal supersymmetric standard model,''
  Prog.\ Theor.\ Phys.\  {\bf 85}, 1 (1991);
  %%CITATION = PTPKA,85,1;%%
  %1124 citations counted in INSPIRE as of 15 Oct 2014
  %\cite{Ellis:1990nz}
%\bibitem{Ellis:1990nz} 
  J.~R.~Ellis, G.~Ridolfi and F.~Zwirner,
  %``Radiative corrections to the masses of supersymmetric Higgs bosons,''
  Phys.\ Lett.\ B {\bf 257}, 83 (1991).
  %%CITATION = PHLTA,B257,83;%%
  %1237 citations counted in INSPIRE as of 15 Oct 2014
  
  %\cite{Haber:1993an}
\bibitem{Haber:1993an} 
  H.~E.~Haber and R.~Hempfling,
  %``The Renormalization group improved Higgs sector of the minimal supersymmetric model,''
  Phys.\ Rev.\ D {\bf 48}, 4280 (1993)
  [hep-ph/9307201].
  %%CITATION = HEP-PH/9307201;%%
  %295 citations counted in INSPIRE as of 15 Oct 2014


\bibitem{mssmhiggsupperbound}
  See e.g.,
%\cite{Carena:2000dp}
%\bibitem{Carena:2000dp} 
  M.~Carena, H.~E.~Haber, S.~Heinemeyer, W.~Hollik, C.~E.~M.~Wagner and G.~Weiglein,
  %``Reconciling the two loop diagrammatic and effective field theory computations of the mass of the lightest CP - even Higgs boson in the MSSM,''
  Nucl.\ Phys.\ B {\bf 580}, 29 (2000)
  [hep-ph/0001002];
  %%CITATION = HEP-PH/0001002;%%
  %303 citations counted in INSPIRE as of 15 Oct 2014
  %\cite{Degrassi:2002fi}
%\bibitem{Degrassi:2002fi} 
  G.~Degrassi, S.~Heinemeyer, W.~Hollik, P.~Slavich and G.~Weiglein,
  %``Towards high precision predictions for the MSSM Higgs sector,''
  Eur.\ Phys.\ J.\ C {\bf 28}, 133 (2003)
  [hep-ph/0212020].
  %%CITATION = HEP-PH/0212020;%%
  %673 citations counted in INSPIRE as of 15 Oct 2014


\bibitem{Giudice:2011cg} 
  G.~F.~Giudice and A.~Strumia,
  %``Probing High-Scale and Split Supersymmetry with Higgs Mass Measurements,''
  Nucl.\ Phys.\ B {\bf 858}, 63 (2012)
  [arXiv:1108.6077 [hep-ph]].
  %%CITATION = ARXIV:1108.6077;%%
  
  %\cite{Bagnaschi:2014rsa}
\bibitem{Bagnaschi:2014rsa} 
  E.~Bagnaschi, G.~F.~Giudice, P.~Slavich and A.~Strumia,
  %``Higgs Mass and Unnatural Supersymmetry,''
  JHEP {\bf 1409}, 092 (2014)
  [arXiv:1407.4081 [hep-ph]].
  %%CITATION = ARXIV:1407.4081;%%
  %7 citations counted in INSPIRE as of 15 Oct 2014

\bibitem{Draper:2013oza} 
  P.~Draper, G.~Lee and C.~E.~M.~Wagner,
  %``Precise estimates of the Higgs mass in heavy supersymmetry,''
  Phys.\ Rev.\ D {\bf 89}, 055023 (2014)
  [arXiv:1312.5743 [hep-ph]].
  %%CITATION = ARXIV:1312.5743;%%

\bibitem{Vega:2015fna} 
  J.~P.~Vega and G.~Villadoro,
  %``SusyHD: Higgs mass Determination in Supersymmetry,''
  JHEP {\bf 1507}, 159 (2015)
  [arXiv:1504.05200 [hep-ph]].
  %%CITATION = ARXIV:1504.05200;%%
 
  
%\cite{Lee:2015uza}
\bibitem{Lee:2015uza} 
  G.~Lee and C.~E.~M.~Wagner,
  %``Higgs bosons in heavy supersymmetry with an intermediate m$_A$,''
  Phys.\ Rev.\ D {\bf 92}, no. 7, 075032 (2015)
  doi:10.1103/PhysRevD.92.075032
  [arXiv:1508.00576 [hep-ph]].
  %%CITATION = doi:10.1103/PhysRevD.92.075032;%%
  %73 citations counted in INSPIRE as of 14 Aug 2019
  
  %\cite{Brust:2011tb}
\bibitem{Brust:2011tb} 
  C.~Brust, A.~Katz, S.~Lawrence and R.~Sundrum,
  %``SUSY, the Third Generation and the LHC,''
  JHEP {\bf 1203}, 103 (2012)
  doi:10.1007/JHEP03(2012)103
  [arXiv:1110.6670 [hep-ph]].
  %%CITATION = doi:10.1007/JHEP03(2012)103;%%
  %333 citations counted in INSPIRE as of 19 Sep 2019  
    
  %\cite{Baum:2017enm}
\bibitem{Baum:2017enm} 
  S.~Baum, M.~Carena, N.~R.~Shah and C.~E.~M.~Wagner,
  %``Higgs portals for thermal Dark Matter. EFT perspectives and the NMSSM,''
  JHEP {\bf 1804}, 069 (2018)
  doi:10.1007/JHEP04(2018)069
  [arXiv:1712.09873 [hep-ph]].
  %%CITATION = doi:10.1007/JHEP04(2018)069;%%
  %19 citations counted in INSPIRE as of 06 Sep 2019 

\bibitem{Ellwanger:2011aa} 
  U.~Ellwanger,
  %``A Higgs boson near 125 GeV with enhanced di-photon signal in the NMSSM,''
  JHEP {\bf 1203}, 044 (2012)
  [arXiv:1112.3548 [hep-ph]].
  %%CITATION = ARXIV:1112.3548;%%
  %211 citations counted in INSPIRE as of 28 Oct 2015
%\cite{Gunion:2012zd}


\bibitem{Gunion:2012zd} 
  J.~F.~Gunion, Y.~Jiang and S.~Kraml,
  %``The Constrained NMSSM and Higgs near 125 GeV,''
  Phys.\ Lett.\ B {\bf 710}, 454 (2012)
  [arXiv:1201.0982 [hep-ph]].
  %%CITATION = ARXIV:1201.0982;%%
  %124 citations counted in INSPIRE as of 28 Oct 2015

\bibitem{King:2012is} 
  S.~F.~King, M.~Muhlleitner and R.~Nevzorov,
  %``NMSSM Higgs Benchmarks Near 125 GeV,''
  Nucl.\ Phys.\ B {\bf 860}, 207 (2012)
  [arXiv:1201.2671 [hep-ph]].
  %%CITATION = ARXIV:1201.2671;%%
  %168 citations counted in INSPIRE as of 28 Oct 2015

\bibitem{Cao:2012fz} 
  J.~J.~Cao, Z.~X.~Heng, J.~M.~Yang, Y.~M.~Zhang and J.~Y.~Zhu,
  %``A SM-like Higgs near 125 GeV in low energy SUSY: a comparative study for MSSM and NMSSM,''
  JHEP {\bf 1203}, 086 (2012)
  [arXiv:1202.5821 [hep-ph]].
  %%CITATION = ARXIV:1202.5821;%%
  %256 citations counted in INSPIRE as of 28 Oct 2015

\bibitem{Vasquez:2012hn} 
  D.~A.~Vasquez, G.~Belanger, C.~Boehm, J.~Da Silva, P.~Richardson and C.~Wymant,
  %``The 125 GeV Higgs in the NMSSM in light of LHC results and astrophysics constraints,''
  Phys.\ Rev.\ D {\bf 86}, 035023 (2012)
  [arXiv:1203.3446 [hep-ph]].
  %%CITATION = ARXIV:1203.3446;%%
  %84 citations counted in INSPIRE as of 28 Oct 2015

\bibitem{Ellwanger:2012ke} 
  U.~Ellwanger and C.~Hugonie,
  %``Higgs bosons near 125 GeV in the NMSSM with constraints at the GUT scale,''
  Adv.\ High Energy Phys.\  {\bf 2012}, 625389 (2012)
  [arXiv:1203.5048 [hep-ph]].
  %%CITATION = ARXIV:1203.5048;%%
  %103 citations counted in INSPIRE as of 28 Oct 2015

\bibitem{Agashe:2012zq} 
  K.~Agashe, Y.~Cui and R.~Franceschini,
  %``Natural Islands for a 125 GeV Higgs in the scale-invariant NMSSM,''
  JHEP {\bf 1302}, 031 (2013)
  [arXiv:1209.2115 [hep-ph]].
  %%CITATION = ARXIV:1209.2115;%%
  %70 citations counted in INSPIRE as of 28 Oct 2015

\bibitem{Kowalska:2012gs} 
  K.~Kowalska, S.~Munir, L.~Roszkowski, E.~M.~Sessolo, S.~Trojanowski and Y.~L.~S.~Tsai,
  %``Constrained next-to-minimal supersymmetric standard model with a 126 GeV Higgs boson: A global analysis,''
  Phys.\ Rev.\ D {\bf 87}, 115010 (2013)
  [arXiv:1211.1693 [hep-ph]].
  %%CITATION = ARXIV:1211.1693;%%
  %55 citations counted in INSPIRE as of 28 Oct 2015

\bibitem{King:2012tr} 
  S.~F.~King, M.~Mühlleitner, R.~Nevzorov and K.~Walz,
  %``Natural NMSSM Higgs Bosons,''
  Nucl.\ Phys.\ B {\bf 870}, 323 (2013)
  [arXiv:1211.5074 [hep-ph]].
  %%CITATION = ARXIV:1211.5074;%%
  %86 citations counted in INSPIRE as of 28 Oct 2015

\bibitem{Gherghetta:2012gb} 
  T.~Gherghetta, B.~von Harling, A.~D.~Medina and M.~A.~Schmidt,
  %``The Scale-Invariant NMSSM and the 126 GeV Higgs Boson,''
  JHEP {\bf 1302}, 032 (2013)
  [arXiv:1212.5243 [hep-ph]].
  %%CITATION = ARXIV:1212.5243;%%
  %85 citations counted in INSPIRE as of 28 Oct 2015

\bibitem{Barbieri:2013hxa} 
  R.~Barbieri, D.~Buttazzo, K.~Kannike, F.~Sala and A.~Tesi,
  %``Exploring the Higgs sector of a most natural NMSSM,''
  Phys.\ Rev.\ D {\bf 87}, no. 11, 115018 (2013)
  [arXiv:1304.3670 [hep-ph]].
  %%CITATION = ARXIV:1304.3670;%%
  %51 citations counted in INSPIRE as of 28 Oct 2015

\bibitem{Badziak:2013bda} 
  M.~Badziak, M.~Olechowski and S.~Pokorski,
  %``New Regions in the NMSSM with a 125 GeV Higgs,''
  JHEP {\bf 1306}, 043 (2013)
  [arXiv:1304.5437 [hep-ph]].
  %%CITATION = ARXIV:1304.5437;%%
  %50 citations counted in INSPIRE as of 28 Oct 2015

\bibitem{Ellwanger:2013ova} 
  U.~Ellwanger,
  %``Higgs pair production in the NMSSM at the LHC,''
  JHEP {\bf 1308}, 077 (2013)
  [arXiv:1306.5541 [hep-ph]].
  %%CITATION = ARXIV:1306.5541;%%
  %52 citations counted in INSPIRE as of 28 Oct 2015

 %\cite{Domingo:2015eea}
\bibitem{Domingo:2015eea} 
  F.~Domingo and G.~Weiglein,
  %``NMSSM interpretations of the observed Higgs signal,''
  arXiv:1509.07283 [hep-ph].
  %%CITATION = ARXIV:1509.07283;%%  

%\cite{King:2014xwa}
\bibitem{King:2014xwa} 
  S.~F.~King, M.~M\"uhlleitner, R.~Nevzorov and K.~Walz,
  %``Discovery Prospects for NMSSM Higgs Bosons at the High-Energy Large Hadron Collider,''
  Phys.\ Rev.\ D {\bf 90}, 095014 (2014)
  [arXiv:1408.1120 [hep-ph]].
  %%CITATION = ARXIV:1408.1120;%%
  %21 citations counted in INSPIRE as of 09 sept. 2015
 
 %\cite{Ellis:1990wk}
\bibitem{Ellis:1990wk} 
  J.~R.~Ellis, S.~Kelley and D.~V.~Nanopoulos,
  %``Probing the desert using gauge coupling unification,''
  Phys.\ Lett.\ B {\bf 260}, 131 (1991).
  doi:10.1016/0370-2693(91)90980-5
  %%CITATION = doi:10.1016/0370-2693(91)90980-5;%%
  %1043 citations counted in INSPIRE as of 26 Nov 2019
   
  %\cite{Ellis:1990wk} 
  %\cite{Langacker:1991an}
\bibitem{Langacker:1991an} 
  P.~Langacker and M.~x.~Luo,
  %``Implications of precision electroweak experiments for $M_t$, $\rho_{0}$, $\sin^2\theta_W$ and grand unification,''
  Phys.\ Rev.\ D {\bf 44}, 817 (1991).
  doi:10.1103/PhysRevD.44.817
  %%CITATION = doi:10.1103/PhysRevD.44.817;%%
  %1426 citations counted in INSPIRE as of 26 Nov 2019
  
    %\cite{Ellis:1990wk} 
  %\cite{Langacker:1991an}
  %\cite{Amaldi:1991cn}
\bibitem{Amaldi:1991cn} 
  U.~Amaldi, W.~de Boer and H.~Furstenau,
  %``Comparison of grand unified theories with electroweak and strong coupling constants measured at LEP,''
  Phys.\ Lett.\ B {\bf 260}, 447 (1991).
  doi:10.1016/0370-2693(91)91641-8
  %%CITATION = doi:10.1016/0370-2693(91)91641-8;%%
  
  
  %\cite{Aaboud:2017ayj}
\bibitem{Aaboud:2017ayj} 
  M.~Aaboud {\it et al.} [ATLAS Collaboration],
  %``Search for a scalar partner of the top quark in the jets plus missing transverse momentum final state at $\sqrt{s}$=13 TeV with the ATLAS detector,''
  JHEP {\bf 1712}, 085 (2017)
  doi:10.1007/JHEP12(2017)085
  [arXiv:1709.04183 [hep-ex]].
  %%CITATION = doi:10.1007/JHEP12(2017)085;%%
  %99 citations counted in INSPIRE as of 23 Nov 2019
  
  %\cite{Aaboud:2017aeu}
\bibitem{Aaboud:2017aeu} 
  M.~Aaboud {\it et al.} [ATLAS Collaboration],
  %``Search for top-squark pair production in final states with one lepton, jets, and missing transverse momentum using 36 fb$^{?1}$ of $ \sqrt{s}=13 $ TeV pp collision data with the ATLAS detector,''
  JHEP {\bf 1806}, 108 (2018)
  doi:10.1007/JHEP06(2018)108
  [arXiv:1711.11520 [hep-ex]].
  %%CITATION = doi:10.1007/JHEP06(2018)108;%%
  %87 citations counted in INSPIRE as of 22 Nov 2019
  
  %\cite{CMS:2019qkm}
\bibitem{CMS:2019qkm} 
  CMS Collaboration [CMS Collaboration],
  %``Search for direct top squark pair production in events with one lepton, jets and missing transverse energy at 13 TeV,''
  CMS-PAS-SUS-19-009.
  %%CITATION = CMS-PAS-SUS-19-009;%%
  %4 citations counted in INSPIRE as of 22 Nov 2019
  
  %\cite{ATLAS:2019oho}
\bibitem{ATLAS:2019oho} 
  The ATLAS collaboration [ATLAS Collaboration],
  %``Search for direct top squark pair production in the 3-body decay mode with a final state containing one lepton, jets, and missing transverse momentum in $\sqrt{s}=13$TeV $pp$ collision data with the ATLAS detector,''
  ATLAS-CONF-2019-017.
  %%CITATION = ATLAS-CONF-2019-017;%%
  %8 citations counted in INSPIRE as of 22 Nov 2019
  
  
 
%\cite{ATLAS:2018doi}
\bibitem{ATLAS:2018doi} 
  The ATLAS collaboration [ATLAS Collaboration],
  %``Combined measurements of Higgs boson production and decay using up to 80 fb$^{-1}$ of proton--proton collision data at $\sqrt{s}=$ 13 TeV collected with the ATLAS experiment,''
  ATLAS-CONF-2018-031.
  %%CITATION = ATLAS-CONF-2018-031;%%
  %28 citations counted in INSPIRE as of 08 Jan 2019 
 
%\cite{Sirunyan:2018koj}
\bibitem{Sirunyan:2018koj} 
  A.~M.~Sirunyan {\it et al.} [CMS Collaboration],
  %``Combined measurements of Higgs boson couplings in proton?proton collisions at $\sqrt{s}=13\,\text {Te}\text {V} $,''
  Eur.\ Phys.\ J.\ C {\bf 79}, no. 5, 421 (2019)
  doi:10.1140/epjc/s10052-019-6909-y
  [arXiv:1809.10733 [hep-ex]].
  %%CITATION = doi:10.1140/epjc/s10052-019-6909-y;%%
  %105 citations counted in INSPIRE as of 14 Aug 2019 
  
  %\cite{Aad:2019mbh}
\bibitem{Aad:2019mbh} 
  G.~Aad {\it et al.} [ATLAS Collaboration],
  %``Combined measurements of Higgs boson production and decay using up to $80$ fb$^{-1}$ of proton-proton collision data at $\sqrt{s}=$ 13 TeV collected with the ATLAS experiment,''
  arXiv:1909.02845 [hep-ex].
  %%CITATION = ARXIV:1909.02845;%%
  %8 citations counted in INSPIRE as of 15 Nov 2019
  
   

 %\cite{Coyle:2018ydo}
\bibitem{Coyle:2018ydo} 
  N.~M.~Coyle, B.~Li and C.~E.~M.~Wagner,
  %``Wrong sign bottom Yukawa coupling in low energy supersymmetry,''
  Phys.\ Rev.\ D {\bf 97}, no. 11, 115028 (2018)
  doi:10.1103/PhysRevD.97.115028
  [arXiv:1802.09122 [hep-ph]].
  %%CITATION = doi:10.1103/PhysRevD.97.115028;%%
  
   %\cite{Haber:1993an}
\bibitem{Haber:1993an} 
  H.~E.~Haber and R.~Hempfling,
  %``The Renormalization group improved Higgs sector of the minimal supersymmetric model,''
  Phys.\ Rev.\ D {\bf 48}, 4280 (1993)
  doi:10.1103/PhysRevD.48.4280
  [hep-ph/9307201].
  %%CITATION = doi:10.1103/PhysRevD.48.4280;%%
  %404 citations counted in INSPIRE as of 01 Dec 2019
  
  %\cite{Haber:1993an}
  %\cite{Carena:1995bx}
\bibitem{Carena:1995bx} 
  M.~Carena, J.~R.~Espinosa, M.~Quiros and C.~E.~M.~Wagner,
  %``Analytical expressions for radiatively corrected Higgs masses and couplings in the MSSM,''
  Phys.\ Lett.\ B {\bf 355}, 209 (1995)
  doi:10.1016/0370-2693(95)00694-G
  [hep-ph/9504316].
  %%CITATION = doi:10.1016/0370-2693(95)00694-G;%%
  %765 citations counted in INSPIRE as of 01 Dec 2019
  



%\cite{Harnik:2003rs}
\bibitem{Harnik:2003rs} 
  R.~Harnik, G.~D.~Kribs, D.~T.~Larson and H.~Murayama,
  %``The Minimal supersymmetric fat Higgs model,''
  Phys.\ Rev.\ D {\bf 70}, 015002 (2004)
  doi:10.1103/PhysRevD.70.015002
  [hep-ph/0311349].
  %%CITATION = doi:10.1103/PhysRevD.70.015002;%%
  %220 citations counted in INSPIRE as of 29 Jul 2019
  
  %\cite{Chang:2004db}
\bibitem{Chang:2004db} 
  S.~Chang, C.~Kilic and R.~Mahbubani,
  %``The New fat Higgs: Slimmer and more attractive,''
  Phys.\ Rev.\ D {\bf 71}, 015003 (2005)
  doi:10.1103/PhysRevD.71.015003
  [hep-ph/0405267].
  %%CITATION = doi:10.1103/PhysRevD.71.015003;%%
  %110 citations counted in INSPIRE as of 14 Aug 2019
  
    %\cite{Chang:2004db}
  %\cite{Delgado:2005fq}
\bibitem{Delgado:2005fq} 
  A.~Delgado and T.~M.~P.~Tait,
  %``A Fat Higgs with a Fat top,''
  JHEP {\bf 0507}, 023 (2005)
  doi:10.1088/1126-6708/2005/07/023
  [hep-ph/0504224].
  %%CITATION = doi:10.1088/1126-6708/2005/07/023;%%
  %82 citations counted in INSPIRE as of 14 Aug 2019
  
  %\cite{Georgi:1984iz}
\bibitem{Georgi:1984iz} 
  H.~Georgi, A.~Manohar and G.~W.~Moore,
  %``Constraints on a Two Higgs Interpretation of the Zeta (8.3),''
  Phys.\ Lett.\  {\bf 149B}, 234 (1984).
  doi:10.1016/0370-2693(84)91591-0
  %%CITATION = doi:10.1016/0370-2693(84)91591-0;%%
  %28 citations counted in INSPIRE as of 28 Aug 2019
  
  %\cite{Georgi:1986kr}
\bibitem{Georgi:1986kr} 
  H.~Georgi and L.~Randall,
  %``Flavor Conserving CP Violation in Invisible Axion Models,''
  Nucl.\ Phys.\ B {\bf 276}, 241 (1986).
  doi:10.1016/0550-3213(86)90022-2
  %%CITATION = doi:10.1016/0550-3213(86)90022-2;%%
  %275 citations counted in INSPIRE as of 28 Aug 2019
  
  %\cite{Luty:1997fk}
\bibitem{Luty:1997fk} 
  M.~A.~Luty,
  %``Naive dimensional analysis and supersymmetry,''
  Phys.\ Rev.\ D {\bf 57}, 1531 (1998)
  doi:10.1103/PhysRevD.57.1531
  [hep-ph/9706235].
  %%CITATION = doi:10.1103/PhysRevD.57.1531;%%
  %170 citations counted in INSPIRE as of 28 Aug 2019
  
  
  %\cite{Cohen:1997rt}
\bibitem{Cohen:1997rt} 
  A.~G.~Cohen, D.~B.~Kaplan and A.~E.~Nelson,
  %``Counting 4 pis in strongly coupled supersymmetry,''
  Phys.\ Lett.\ B {\bf 412}, 301 (1997)
  doi:10.1016/S0370-2693(97)00995-7
  [hep-ph/9706275].
  %%CITATION = doi:10.1016/S0370-2693(97)00995-7;%%
  %183 citations counted in INSPIRE as of 28 Aug 2019

    
  %\cite{Giudice:1988yz}
\bibitem{Giudice:1988yz} 
  G.~F.~Giudice and A.~Masiero,
  %``A Natural Solution to the mu Problem in Supergravity Theories,''
  Phys.\ Lett.\ B {\bf 206}, 480 (1988).
  doi:10.1016/0370-2693(88)91613-9
  %%CITATION = doi:10.1016/0370-2693(88)91613-9;%%
  %946 citations counted in INSPIRE as of 12 Aug 2019
  
    %\cite{Giudice:1997ni}
\bibitem{Giudice:1997ni} 
  G.~F.~Giudice and R.~Rattazzi,
  %``Extracting supersymmetry breaking effects from wave function renormalization,''
  Nucl.\ Phys.\ B {\bf 511}, 25 (1998)
  doi:10.1016/S0550-3213(97)00647-0
  [hep-ph/9706540].
  %%CITATION = doi:10.1016/S0550-3213(97)00647-0;%%
  %317 citations counted in INSPIRE as of 22 Oct 2019
  

%\cite{Dicus:1994bm}
\bibitem{Dicus:1994bm} 
  D.~Dicus, A.~Stange and S.~Willenbrock,
  %``Higgs decay to top quarks at hadron colliders,''
  Phys.\ Lett.\ B {\bf 333}, 126 (1994)
  doi:10.1016/0370-2693(94)91017-0
  [hep-ph/9404359].
  %%CITATION = doi:10.1016/0370-2693(94)91017-0;%%

%\cite{Craig:2015jba}
\bibitem{Craig:2015jba} 
  N.~Craig, F.~D'Eramo, P.~Draper, S.~Thomas and H.~Zhang,
  %``The Hunt for the Rest of the Higgs Bosons,''
  JHEP {\bf 1506}, 137 (2015)
  doi:10.1007/JHEP06(2015)137
  [arXiv:1504.04630 [hep-ph]].
  %%CITATION = doi:10.1007/JHEP06(2015)137;%%
  
%\cite{Jung:2015gta}
\bibitem{Jung:2015gta} 
  S.~Jung, J.~Song and Y.~W.~Yoon,
  %``Dip or nothingness of a Higgs resonance from the interference with a complex phase,''
  Phys.\ Rev.\ D {\bf 92}, no. 5, 055009 (2015)
  doi:10.1103/PhysRevD.92.055009
  [arXiv:1505.00291 [hep-ph]].
  %%CITATION = doi:10.1103/PhysRevD.92.055009;%%



  %\cite{Gori:2016zto}
\bibitem{Gori:2016zto} 
  S.~Gori, I.~W.~Kim, N.~R.~Shah and K.~M.~Zurek,
  %``Closing the Wedge: Search Strategies for Extended Higgs Sectors with Heavy Flavor Final States,''
  Phys.\ Rev.\ D {\bf 93}, no. 7, 075038 (2016)
  doi:10.1103/PhysRevD.93.075038
  [arXiv:1602.02782 [hep-ph]].
  %%CITATION = doi:10.1103/PhysRevD.93.075038;%%

%\cite{Dicus:1994bm}
%\cite{Carena:2016npr}
\bibitem{Carena:2016npr} 
  M.~Carena and Z.~Liu,
  %``Challenges and opportunities for heavy scalar searches in the $ t\overline{t} $ channel at the LHC,''
  JHEP {\bf 1611}, 159 (2016)
  doi:10.1007/JHEP11(2016)159
  [arXiv:1608.07282 [hep-ph]].
  %%CITATION = doi:10.1007/JHEP11(2016)159;%%
  
  
  %\cite{Aaboud:2017sjh}
\bibitem{Aaboud:2017sjh} 
  M.~Aaboud {\it et al.} [ATLAS Collaboration],
  %``Search for additional heavy neutral Higgs and gauge bosons in the ditau final state produced in 36 fb$^{?1}$ of pp collisions at $ \sqrt{s}=13 $ TeV with the ATLAS detector,''
  JHEP {\bf 1801}, 055 (2018)
  doi:10.1007/JHEP01(2018)055
  [arXiv:1709.07242 [hep-ex]].
  %%CITATION = doi:10.1007/JHEP01(2018)055;%%

%\cite{Aaboud:2017sjh}
%\cite{Sirunyan:2018zut}
\bibitem{Sirunyan:2018zut} 
  A.~M.~Sirunyan {\it et al.} [CMS Collaboration],
  %``Search for additional neutral MSSM Higgs bosons in the $\tau\tau$ final state in proton-proton collisions at $\sqrt{s}=$ 13 TeV,''
  JHEP {\bf 1809}, 007 (2018)
  doi:10.1007/JHEP09(2018)007
  [arXiv:1803.06553 [hep-ex]].
  %%CITATION = doi:10.1007/JHEP09(2018)007;%%
 

%\cite{Aaboud:2017sjh}
%\cite{Sirunyan:2018zut}
 %\cite{Bahl:2018zmf}
\bibitem{Bahl:2018zmf} 
  E.~Bagnaschi {\it et al.},
  %``MSSM Higgs Boson Searches at the LHC: Benchmark Scenarios for Run 2 and Beyond,''
  Eur.\ Phys.\ J.\ C {\bf 79}, no. 7, 617 (2019)
  doi:10.1140/epjc/s10052-019-7114-8
  [arXiv:1808.07542 [hep-ph]].
  %%CITATION = doi:10.1140/epjc/s10052-019-7114-8;%%
 




%\cite{Ellwanger:2004xm}
\bibitem{Ellwanger:2004xm}
  U.~Ellwanger, J.~F.~Gunion and C.~Hugonie,
  %``NMHDECAY: A Fortran code for the Higgs masses, couplings and decay widths in the NMSSM,''
  JHEP {\bf 0502}, 066 (2005)
  doi:10.1088/1126-6708/2005/02/066
  [hep-ph/0406215].
  %%CITATION = doi:10.1088/1126-6708/2005/02/066;%%
  %335 citations counted in INSPIRE as of 21 Nov 2016


%\cite{Hewett:1992is,Misiak:2015xwa,Misiak:2017zan,Misiak:2017bgg}
%\cite{Hewett:1992is}
\bibitem{Hewett:1992is} 
  J.~L.~Hewett,
  %``Can b ---> s gamma close the supersymmetric Higgs production window?,''
  Phys.\ Rev.\ Lett.\  {\bf 70}, 1045 (1993)
  doi:10.1103/PhysRevLett.70.1045
  [hep-ph/9211256].
  %%CITATION = doi:10.1103/PhysRevLett.70.1045;%%
  %234 citations counted in INSPIRE as of 28 Feb 2020

  
  %\cite{Misiak:2015xwa}
\bibitem{Misiak:2015xwa} 
  M.~Misiak {\it et al.},
  %``Updated NNLO QCD predictions for the weak radiative B-meson decays,''
  Phys.\ Rev.\ Lett.\  {\bf 114}, no. 22, 221801 (2015)
  doi:10.1103/PhysRevLett.114.221801
  [arXiv:1503.01789 [hep-ph]].
  %%CITATION = doi:10.1103/PhysRevLett.114.221801;%%
  %289 citations counted in INSPIRE as of 28 Feb 2020
  
  %\cite{Misiak:2017zan}
\bibitem{Misiak:2017zan} 
  M.~Misiak,
  %``Bounds on $M_{H^\pm }$ from $\bar B \to X_{s,d}\gamma $ Decays,''
  Acta Phys.\ Polon.\ B {\bf 48}, 2173 (2017).
  doi:10.5506/APhysPolB.48.2173
  %%CITATION = doi:10.5506/APhysPolB.48.2173;%%
  %1 citations counted in INSPIRE as of 28 Feb 2020
  
  %\cite{Misiak:2017bgg}
\bibitem{Misiak:2017bgg} 
  M.~Misiak and M.~Steinhauser,
  %``Weak radiative decays of the B meson and bounds on $M_{H^\pm }$ in the Two-Higgs-Doublet Model,''
  Eur.\ Phys.\ J.\ C {\bf 77}, no. 3, 201 (2017)
  doi:10.1140/epjc/s10052-017-4776-y
  [arXiv:1702.04571 [hep-ph]].
  %%CITATION = doi:10.1140/epjc/s10052-017-4776-y;%%
  %142 citations counted in INSPIRE as of 28 Feb 2020


%\cite{Barbieri:1993av,Degrassi:2000qf,Carena:2000uj,Buras:2002vd}
%\cite{Barbieri:1993av}
\bibitem{Barbieri:1993av} 
  R.~Barbieri and G.~F.~Giudice,
  %``b ---> s gamma decay and supersymmetry,''
  Phys.\ Lett.\ B {\bf 309}, 86 (1993)
  doi:10.1016/0370-2693(93)91508-K
  [hep-ph/9303270].
  %%CITATION = doi:10.1016/0370-2693(93)91508-K;%%
  %374 citations counted in INSPIRE as of 28 Feb 2020
  
%\cite{Degrassi:2000qf}
\bibitem{Degrassi:2000qf} 
  G.~Degrassi, P.~Gambino and G.~F.~Giudice,
  %``B ---> X(s gamma) in supersymmetry: Large contributions beyond the leading order,''
  JHEP {\bf 0012}, 009 (2000)
  doi:10.1088/1126-6708/2000/12/009
  [hep-ph/0009337].
  %%CITATION = doi:10.1088/1126-6708/2000/12/009;%%
  %431 citations counted in INSPIRE as of 28 Feb 2020
  
  %\cite{Carena:2000uj}
\bibitem{Carena:2000uj} 
  M.~Carena, D.~Garcia, U.~Nierste and C.~E.~M.~Wagner,
  %``$b \to s \gamma$ and supersymmetry with large $\tan\beta$,''
  Phys.\ Lett.\ B {\bf 499}, 141 (2001)
  doi:10.1016/S0370-2693(01)00009-0
  [hep-ph/0010003].
  %%CITATION = doi:10.1016/S0370-2693(01)00009-0;%%
  %397 citations counted in INSPIRE as of 28 Feb 2020

%\cite{Buras:2002vd}
\bibitem{Buras:2002vd} 
  A.~J.~Buras, P.~H.~Chankowski, J.~Rosiek and L.~Slawianowska,
  %``$\Delta M_{d,s}, B^0{d,s} \to \mu^{+} \mu^{-}$ and $B \to X_{s} \gamma$ in supersymmetry at large $\tan\beta$,''
  Nucl.\ Phys.\ B {\bf 659}, 3 (2003)
  doi:10.1016/S0550-3213(03)00190-1
  [hep-ph/0210145].
  %%CITATION = doi:10.1016/S0550-3213(03)00190-1;%%
  %355 citations counted in INSPIRE as of 28 Feb 2020


%\cite{Gabbiani:1996hi}
\bibitem{Gabbiani:1996hi} 
  F.~Gabbiani, E.~Gabrielli, A.~Masiero and L.~Silvestrini,
  %``A Complete analysis of FCNC and CP constraints in general SUSY extensions of the standard model,''
  Nucl.\ Phys.\ B {\bf 477}, 321 (1996)
  doi:10.1016/0550-3213(96)00390-2
  [hep-ph/9604387].
  %%CITATION = doi:10.1016/0550-3213(96)00390-2;%%
  %1353 citations counted in INSPIRE as of 28 Feb 2020
  

    
  %\cite{Arason:1991hu}
\bibitem{Arason:1991hu} 
  H.~Arason, D.~Castano, B.~Keszthelyi, S.~Mikaelian, E.~Piard, P.~Ramond and B.~Wright,
  %``Top quark and Higgs mass bounds from a numerical study of superGUTs,''
  Phys.\ Rev.\ Lett.\  {\bf 67}, 2933 (1991).
  doi:10.1103/PhysRevLett.67.2933
  %%CITATION = doi:10.1103/PhysRevLett.67.2933;%%
  %229 citations counted in INSPIRE as of 12 Aug 2019
  
  %\cite{Carena:1993ag}
\bibitem{Carena:1993ag} 
  M.~Carena, S.~Pokorski and C.~E.~M.~Wagner,
  %``On the unification of couplings in the minimal supersymmetric Standard Model,''
  Nucl.\ Phys.\ B {\bf 406}, 59 (1993)
  doi:10.1016/0550-3213(93)90161-H
  [hep-ph/9303202].
  %%CITATION = doi:10.1016/0550-3213(93)90161-H;%%
  %350 citations counted in INSPIRE as of 26 Nov 2019
  
  %\cite{Arason:1991hu}
  %\cite{Bardeen:1993rv}
\bibitem{Bardeen:1993rv} 
  W.~A.~Bardeen, M.~Carena, S.~Pokorski and C.~E.~M.~Wagner,
  %``Infrared Fixed Point Solution for the Top Quark Mass and Unification of Couplings in the MSSM,''
  Phys.\ Lett.\ B {\bf 320}, 110 (1994)
  doi:10.1016/0370-2693(94)90832-X
  [hep-ph/9309293].
  %%CITATION = doi:10.1016/0370-2693(94)90832-X;%%
  %121 citations counted in INSPIRE as of 12 Aug 2019
    
  %\cite{Allanach:1994zd}
\bibitem{Allanach:1994zd} 
  B.~C.~Allanach and S.~F.~King,
  %``Bottom - tau Yukawa unification in the next-to-minimal supersymmetric Standard Model,''
  Phys.\ Lett.\ B {\bf 328}, 360 (1994)
  doi:10.1016/0370-2693(94)91491-5
  [hep-ph/9403212].
  %%CITATION = doi:10.1016/0370-2693(94)91491-5;%%
  %28 citations counted in INSPIRE as of 14 Aug 2019
  
  %\cite{Carena:1993ag}
\bibitem{Carena:1993ag} 
  M.~Carena, S.~Pokorski and C.~E.~M.~Wagner,
  %``On the unification of couplings in the minimal supersymmetric Standard Model,''
  Nucl.\ Phys.\ B {\bf 406}, 59 (1993)
  doi:10.1016/0550-3213(93)90161-H
  [hep-ph/9303202].
  %%CITATION = doi:10.1016/0550-3213(93)90161-H;%%
  %348 citations counted in INSPIRE as of 14 Aug 2019
  
  %\cite{Hempfling:1993kv}
\bibitem{Hempfling:1993kv} 
  R.~Hempfling,
  %``Yukawa coupling unification with supersymmetric threshold corrections,''
  Phys.\ Rev.\ D {\bf 49}, 6168 (1994).
  doi:10.1103/PhysRevD.49.6168
  %%CITATION = doi:10.1103/PhysRevD.49.6168;%%
  %472 citations counted in INSPIRE as of 14 Aug 2019
  
  %\cite{Hall:1993gn}
\bibitem{Hall:1993gn} 
  L.~J.~Hall, R.~Rattazzi and U.~Sarid,
  %``The Top quark mass in supersymmetric SO(10) unification,''
  Phys.\ Rev.\ D {\bf 50}, 7048 (1994)
  doi:10.1103/PhysRevD.50.7048
  [hep-ph/9306309].
  %%CITATION = doi:10.1103/PhysRevD.50.7048;%%
  %950 citations counted in INSPIRE as of 14 Aug 2019
  
  %\cite{Carena:1994bv}
\bibitem{Carena:1994bv} 
  M.~Carena, M.~Olechowski, S.~Pokorski and C.~E.~M.~Wagner,
  %``Electroweak symmetry breaking and bottom - top Yukawa unification,''
  Nucl.\ Phys.\ B {\bf 426}, 269 (1994)
  doi:10.1016/0550-3213(94)90313-1
  [hep-ph/9402253].
  %%CITATION = doi:10.1016/0550-3213(94)90313-1;%%
  %811 citations counted in INSPIRE as of 14 Aug 2019
  
  %\cite{Langacker:1993xb}
\bibitem{Langacker:1993xb} 
  P.~Langacker and N.~Polonsky,
  %``The Bottom mass prediction in supersymmetric grand unification: Uncertainties and constraints,''
  Phys.\ Rev.\ D {\bf 49}, 1454 (1994)
  doi:10.1103/PhysRevD.49.1454
  [hep-ph/9306205].
  %%CITATION = doi:10.1103/PhysRevD.49.1454;%%
  %216 citations counted in INSPIRE as of 14 Aug 2019
  
  %\cite{Langacker:1994bc}
\bibitem{Langacker:1994bc} 
  P.~Langacker and N.~Polonsky,
  %``Implications of Yukawa unification for the Higgs sector in supersymmetric grand unified models,''
  Phys.\ Rev.\ D {\bf 50}, 2199 (1994)
  doi:10.1103/PhysRevD.50.2199
  [hep-ph/9403306].
  %%CITATION = doi:10.1103/PhysRevD.50.2199;%%
  %135 citations counted in INSPIRE as of 14 Aug 2019
  
  
  %\cite{Schrempp:1994xn}
\bibitem{Schrempp:1994xn} 
  B.~Schrempp,
  %``Infrared fixed points and fixed lines in the top bottom $\tau$ sector in supersymmetric grand unification,''
  Phys.\ Lett.\ B {\bf 344}, 193 (1995)
  doi:10.1016/0370-2693(94)01559-U
  [hep-ph/9411241].
  %%CITATION = doi:10.1016/0370-2693(94)01559-U;%%
  %38 citations counted in INSPIRE as of 14 Aug 2019
  
  %\cite{Kolda:1994ab}
\bibitem{Kolda:1994ab} 
  C.~F.~Kolda, L.~Roszkowski, J.~D.~Wells and G.~L.~Kane,
  %``Predictions for constrained minimal supersymmetry with bottom tau mass unification,''
  Phys.\ Rev.\ D {\bf 50}, 3498 (1994)
  doi:10.1103/PhysRevD.50.3498
  [hep-ph/9404253].
  %%CITATION = doi:10.1103/PhysRevD.50.3498;%%
  %45 citations counted in INSPIRE as of 14 Aug 2019
    
  
  %\cite{Bardeen:1989ds}
\bibitem{Bardeen:1989ds} 
  W.~A.~Bardeen, C.~T.~Hill and M.~Lindner,
  %``Minimal Dynamical Symmetry Breaking of the Standard Model,''
  Phys.\ Rev.\ D {\bf 41}, 1647 (1990).
  doi:10.1103/PhysRevD.41.1647
  %%CITATION = doi:10.1103/PhysRevD.41.1647;%%
  %1156 citations counted in INSPIRE as of 12 Aug 2019
  
  %\cite{Clark:1989tq}
\bibitem{Clark:1989tq} 
  T.~E.~Clark, S.~T.~Love and W.~A.~Bardeen,
  %``The Top Quark Mass in a Supersymmetric Standard Model with Dynamical Symmetry Breaking,''
  Phys.\ Lett.\ B {\bf 237}, 235 (1990).
  doi:10.1016/0370-2693(90)91435-E
  %%CITATION = doi:10.1016/0370-2693(90)91435-E;%%
  %95 citations counted in INSPIRE as of 12 Aug 2019
  
  %\cite{Bardeen:1989ds}
    %\cite{Clark:1989tq}
  %\cite{Carena:1991ky}
\bibitem{Carena:1991ky} 
  M.~Carena, T.~E.~Clark, C.~E.~M.~Wagner, W.~A.~Bardeen and K.~Sasaki,
  %``Dynamical symmetry breaking and the top quark mass in the minimal supersymmetric standard model,''
  Nucl.\ Phys.\ B {\bf 369}, 33 (1992).
  doi:10.1016/0550-3213(92)90377-N
  %%CITATION = doi:10.1016/0550-3213(92)90377-N;%%
  %110 citations counted in INSPIRE as of 12 Aug 2019
   
  
  %\cite{Komine:2001rm}
\bibitem{Komine:2001rm} 
  S.~Komine and M.~Yamaguchi,
  %``Bottom tau unification in SUSY SU(5) GUT and constraints from b ---> s gamma and muon g-2,''
  Phys.\ Rev.\ D {\bf 65}, 075013 (2002)
  doi:10.1103/PhysRevD.65.075013
  [hep-ph/0110032].
  %%CITATION = doi:10.1103/PhysRevD.65.075013;%%
  %54 citations counted in INSPIRE as of 14 Aug 2019
  

%\cite{Sato:2000zh}
\bibitem{Sato:2000zh} 
  J.~Sato, K.~Tobe and T.~Yanagida,
  %``A Constraint on Yukawa coupling unification from lepton flavor violating processes,''
  Phys.\ Lett.\ B {\bf 498}, 189 (2001)
  doi:10.1016/S0370-2693(00)01395-2
  [hep-ph/0010348].
  %%CITATION = doi:10.1016/S0370-2693(00)01395-2;%%
  %81 citations counted in INSPIRE as of 14 Aug 2019  
  
  %\cite{Balazs:2003mm}
\bibitem{Balazs:2003mm} 
  C.~Balazs and R.~Dermisek,
  %``Yukawa coupling unification and nonuniversal gaugino mediation of supersymmetry breaking,''
  JHEP {\bf 0306}, 024 (2003)
  doi:10.1088/1126-6708/2003/06/024
  [hep-ph/0303161].
  %%CITATION = doi:10.1088/1126-6708/2003/06/024;%%
  %47 citations counted in INSPIRE as of 14 Aug 2019
  
  %\cite{Auto:2003ys}
\bibitem{Auto:2003ys} 
  D.~Auto, H.~Baer, C.~Balazs, A.~Belyaev, J.~Ferrandis and X.~Tata,
  %``Yukawa coupling unification in supersymmetric models,''
  JHEP {\bf 0306}, 023 (2003)
  doi:10.1088/1126-6708/2003/06/023
  [hep-ph/0302155].
  %%CITATION = doi:10.1088/1126-6708/2003/06/023;%%
  %135 citations counted in INSPIRE as of 14 Aug 2019
  
  %\cite{Baer:2012by}
\bibitem{Baer:2012by} 
  H.~Baer, I.~Gogoladze, A.~Mustafayev, S.~Raza and Q.~Shafi,
  %``Sparticle mass spectra from SU(5) SUSY GUT models with $b-\tau$ Yukawa coupling unification,''
  JHEP {\bf 1203}, 047 (2012)
  doi:10.1007/JHEP03(2012)047
  [arXiv:1201.4412 [hep-ph]].
  %%CITATION = doi:10.1007/JHEP03(2012)047;%%
  %24 citations counted in INSPIRE as of 14 Aug 2019
  
  %\cite{Branco:2011iw}
\bibitem{Branco:2011iw} 
  G.~C.~Branco, P.~M.~Ferreira, L.~Lavoura, M.~N.~Rebelo, M.~Sher and J.~P.~Silva,
  %``Theory and phenomenology of two-Higgs-doublet models,''
  Phys.\ Rept.\  {\bf 516}, 1 (2012)
  doi:10.1016/j.physrep.2012.02.002
  [arXiv:1106.0034 [hep-ph]].
  %%CITATION = doi:10.1016/j.physrep.2012.02.002;%%
  
 
 


    
\end{thebibliography}
\end{document}